\documentclass[iop]{emulateapj}
\usepackage{times,txfonts,graphicx,aas_macros,natbib,xspace,color,bm}
\pdfoutput=1
\usepackage{ulem}

\renewcommand\emph[1]{\textit{#1}}

\newcommand\tms{\!\times\!}
\newcommand\cdt{\!\cdot\!}

\newcommand\xx{\hat{{\mathbf x}}}
\newcommand\yy{\hat{{\mathbf y}}}
\newcommand\zz{\hat{{\mathbf z}}}

\newcommand\V{\mathbf v}
\newcommand\B{\mathbf B}
\newcommand\U{\mathbf{u}}
\newcommand\mU{\mn{\mathbf{u}}}
\newcommand\mB{\mn{\mathbf{B}}}

\newcommand\EMF{\mbox{\boldmath{${\cal E}$}}}
\newcommand{\OO}{\bm{\Omega}}

\newcommand\betap{\beta_{\rm p}}

\newcommand\tauc{\tau_{\rm c}}

\newcommand{\mn}[1]{\overline{#1}}

\newcommand{\simgt}%
           {\,\hbox{\lower0.35ex\hbox{$\sim$}\llap{\raise0.35ex\hbox{$>$}}}\,}
\newcommand{\simlt}%
           {\,\hbox{\lower0.35ex\hbox{$\sim$}\llap{\raise0.35ex\hbox{$<$}}}\,}

\newcommand\NIII{\textsc{nirvana-iii}\xspace}

\begin{document}

\title{Characterizing the mean-field dynamo in turbulent accretion disks}

\author{
  Oliver~Gressel and Martin E. Pessah
}

\affil{
  Niels Bohr International Academy, The Niels Bohr Institute,
  Blegdamsvej 17, DK-2100, Copenhagen \O, Denmark\\
}

\email{
  gressel@nbi.ku.dk (OG); mpessah@nbi.ku.dk (MP)
}


\begin{abstract}
The formation and evolution of a wide class of astrophysical objects is governed by turbulent, magnetized accretion disks. Understanding their secular dynamics is of primary importance. Apart from enabling mass accretion via the transport of angular momentum, the turbulence affects the long-term evolution of the embedded magnetic flux, which in turn regulates the efficiency of the transport.
In this paper, we take a comprehensive next step towards an effective mean-field model for turbulent astrophysical disks by systematically studying the key properties of magnetorotational turbulence in vertically-stratified, isothermal shearing boxes. This allows us to infer emergent properties of the ensuing chaotic flow as a function of the shear parameter as well as the amount of net-vertical flux.
Using the test-field method, we furthermore characterize the mean-field dynamo coefficients that describe the long-term evolution of large-scale fields. We simultaneously infer the vertical shape and the spectral scale dependence of these closure coefficients, with the latter describing non-local contributions to the turbulent electromotive force. Based on this, we infer a scale-separation ratio of about ten for the large-scale dynamo. We finally determine scaling properties of the mean-field dynamo coefficients. The relevant component of the dynamo $\alpha$~effect is found to scale linearly with the shear rate, as is the corresponding turbulent diffusion, $\eta$. Together, these scalings allow us to predict, in a quantitative manner, the cycle period of the well-known butterfly diagram.
This lends new support to the importance of the $\alpha\Omega$ mechanism in determining the evolution of large-scale magnetic fields in turbulent accretion disks.
\end{abstract}



\section{Introduction}
\label{sec:intro}

Accretion disks are of central importance to the formation and evolution of a wide range of astrophysical objects. If sufficiently ionized, such disks are  subject to the magnetorotational instability (MRI), driving turbulence within the disk body. Since the pioneering work of \citet{1991ApJ...376..214B}, numerical simulations have played a key role in establishing our current knowledge about the non-linearly saturated, turbulent state \citep[see][for a review]{2003LNP...614..329B} -- yet there remain substantial open questions regarding the saturation amplitude of the turbulent stresses, and their scaling with various input parameters.

Ideally, one would strive to simulate the accretion disk in its entirety, but the numerical requirements to treat \emph{turbulent} magnetized disks globally are immense even by today's standards \citep[see][for an estimate]{2003MNRAS.340..969O}. Both the grid resolution \citep{2013ApJ...772..102H} and the domain size \citep{2012ApJ...744..144F} have to be sufficient for obtaining converged results, and existing global models \citep{2000ApJ...528..462H,2009A&A...496..597F,2010A&A...515A..70D,2011MNRAS.416..361B} all have to make compromises of one sort or another.

The need for global simulations (necessary to properly describe the disk wind and jet) on one hand, and the difficulty to resolve all relevant length scales (necessary to obtain converged results for the turbulence) on the other hand, strongly promotes the development of a so-called sub-grid-scale model \citep[in the vein of the large-eddy simulations pioneered by][]{1963MWRv...91...99S}. Such an approach can be regarded as mandatory when integration over long dynamical timescales is required, for instance to follow the secular evolution of the accretion disk, or the stability and symmetry of a protostellar jet \citep{2013ApJ...774...12F}.

Here we advocate a strategy where adequately resolved local shearing-box models with varying shear rate, $q$, and net-vertical flux (NVF) are employed to derive an \emph{effective} parametrization of MRI-generated turbulence with the prospect of later inserting this into global simulations as a sub-grid-scale model. Early phenomenological descriptions for the turbulent stresses \citep{1995PASJ...47..629K,2003MNRAS.340..969O} included the effects of the background shear but ignored any large-scale fields. These phenomenological approaches typically involve a certain number of unconstrained order-unity coefficients. \citet{2009AN....330...92L} and \citet{2012AN....333...78S} have made first attempts to restrict various closure coefficients in heuristic approaches, such as the \citet{2003MNRAS.340..969O} stress model, from direct simulations, pioneering an important future route. \citet{2013MNRAS.428.2255P} have tested such shear-stress relations against global relativistic simulations, and derive a prescription where the turbulent stress depends on radius. Starting from the early phenomenological models, \citet*{2006PhRvL..97v1103P} included for the first time a driving term akin to the linear MRI assuming a constant net-vertical flux. Notably, this simple model already produces a scaling of the Maxwell-to-Reynolds stress ratio according to $\,(4-q)/q\,$ as a function of the dimensionless shear parameter \citep*{2006MNRAS.372..183P}, as well as the total stress as a function of shear \citep*{2008MNRAS.383..683P}. The former relation had previously been found in unstratified MRI simulations by \citet*{1999ApJ...518..394H}. This correspondence may indicate that the linear modes remain important even when affected by turbulent dissipation \citep{2008ApJ...685..406W,2014A&A...567A.139V}. Addressing questions related to the transport properties of the emerging turbulent flow, \citet{2015MNRAS.446.2102N} have recently revisited the subject of shear-rate dependence in MRI turbulence. The authors confirm the scaling of the turbulent stresses with the shear-to-vorticity ratio $\,q/(2-q)\,$, first suggested by \citet*{1996MNRAS.281L..21A}, as well as the relation of \citet*{2006MNRAS.372..183P} quoted above.

While the mentioned approaches primarily focus on the momentum equation, ideally a closure model should also encapsulate the effects of small-scale turbulence (\emph{via} the induction equation) onto the long-term evolution of the large-scale field threading the accretion disk \citep{2010AN....331..101B}.
The relative importance of the turbulent diffusivity versus the turbulent viscosity has previously been assessed in unstratified MRI simulations \citep{2009ApJ...697.1901G,2009A&A...504..309L,2009A&A...507...19F}. It is crucial to appreciate that a large-scale field acts as a net flux locally \citep{2010ApJ...712.1241S} and thus determines the driving of turbulence via linear MRI -- in turn affecting the local value of turbulent transport coefficients governing the redistribution of angular momentum \citep{2012PhyS...86e8202B}. Beyond linear MRI, the variability in disk coronae may be a consequence of the emergence of large-scale azimuthal field \citep{2009ApJ...704L.113B}, and the long-term radial evolution of the net-vertical flux \citep{2012MNRAS.424.2097G,2013MNRAS.430..822G,2014MNRAS.441..852G} may have important implications for the potential driving of a disk wind \citep[see discussion in][]{1994MNRAS.267..235L}, although \citet{2003A&A...398..825V} had found that a sufficiently strong dynamo can support outflows even in the absence of significant vertical fields. The importance of mean-field effects is now also witnessed in global simulations \citep{2012ApJ...744..144F,2014ApJ...784..169J} that show the same cyclic behavior of the azimuthally-averaged magnetic fields that had previously been observed in local boxes only \citep[see discussion in][]{2010MNRAS.405...41G}. Related to this, \citet{2014ApJ...796...29S} have recently argued that the quasi-periodic character of the disk dynamo (which they trigger by hand) may be reflected in episodic ejection events, which can be observed as knots along the jet axis.

As with the turbulent stresses, a promising route to nurturing our insight is to understand the parameter dependence of the various transport coefficients. Initiated by the numerical result of \citet{2008PhRvL.100r4501Y}, who found an exponential amplification of large-scale fields in unstratified shear flow, the role of shear in generating correlations in the mean electromotive force has received renewed interest \citep{2011PhRvL.107y5004H,2012MNRAS.420.2170M,2013JFM...717..395H}. The shear-rate dependence of the dynamo $\alpha$~effect in MRI turbulence was first studied by \citet{2001A&A...378..668Z}, who inferred a (pseudo-) scalar $\alpha$ simply as the ratio of the electromotive force (EMF) to the mean field -- hence ignoring any contributions via a tensorial $\alpha$~effect, and/or the turbulent diffusivity, $\eta$ \citep[also see][]{1995ApJ...446..741B,2000A&A...356.1141Z}. The numerical study of \citeauthor{2001A&A...378..668Z} was prompted by the theoretical work of \citet{2000A&A...362..756R}, who predicted that the \emph{negative} $\alpha$~effect \citep{1998tbha.conf...61B} would only exist in a narrow range of the shear parameter. We are able to confirm this prediction, but also show that this ``buoyant'' $\alpha$~effect is probably \emph{not} responsible for the cyclic patterns seen in space-time diagrams of the mean horizontal magnetic field.  For the \emph{positive} $\alpha$~effect, however, we find a linear dependence on shear, which nicely explains the cycle period that we infer from our simulations as a function of the shear rate -- albeit \emph{not} the direction of propagation. The result nevertheless lends new support to the classical $\alpha\Omega$ dynamo as a natural explanation of the magnetic cycles.

\begin{figure}
  \center\includegraphics[width=0.925\columnwidth]{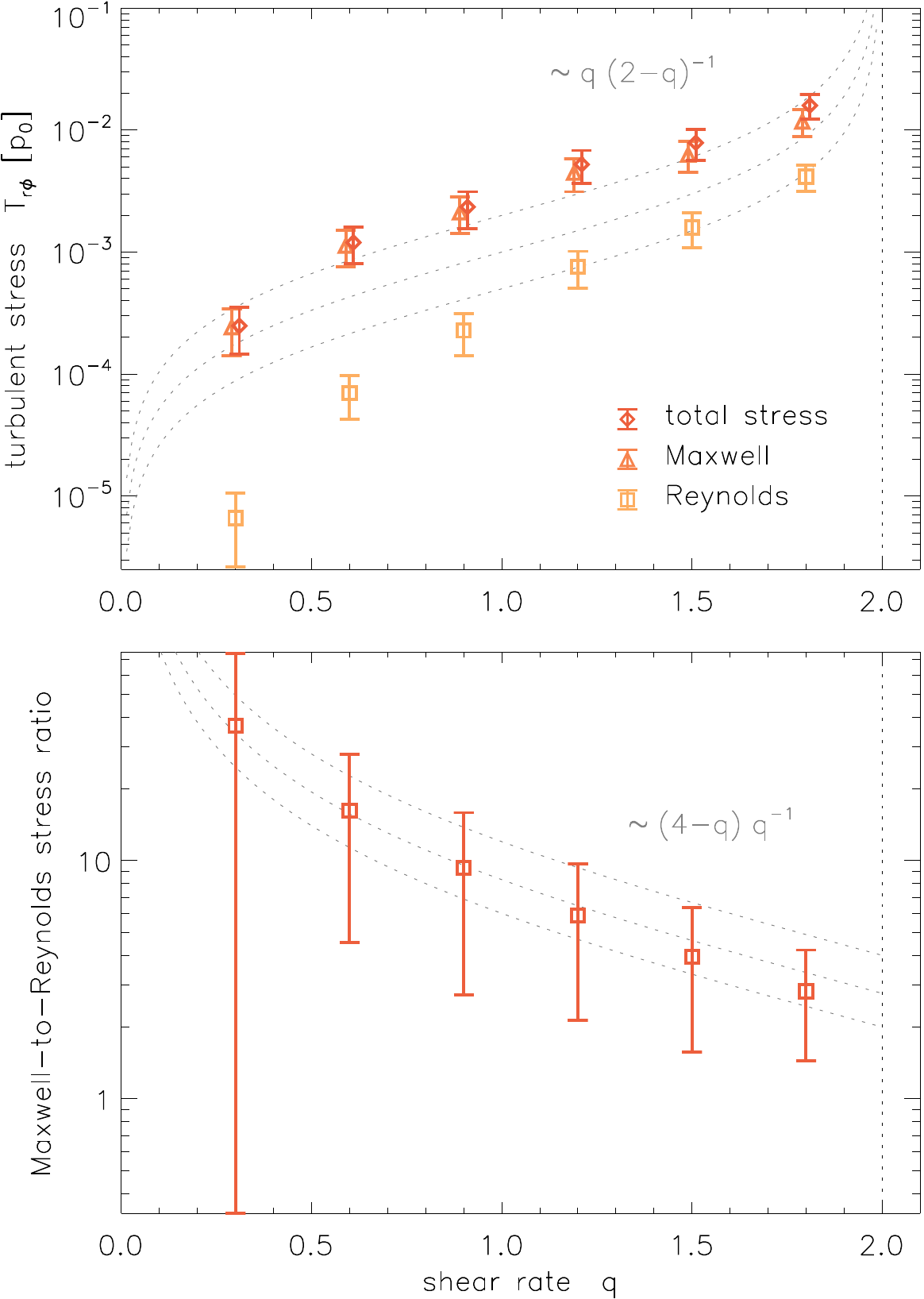}
  \caption{\emph{Top:} Shear-rate dependence of the Reynolds stress (squares), Maxwell stress (triangles), and their sum (diamonds). The dotted lines indicate the scaling with the shear-to-vorticity ratio, $q/(2-q)$ first suggested by \citet{1996MNRAS.281L..21A}. \emph{Bottom:} shear-rate dependence of the Maxwell-to-Reynolds stress ratio. The dotted lines indicate scaling, $(4-q)/q$, introduced by \citet*{2006MNRAS.372..183P}.}
  \label{fig:q_stress}
\end{figure}

The second theme of our investigation is the NVF-dependence of (i) the turbulent stress and (ii) the dynamo coefficients. In contrast to unstratified NVF simulations \citep[that find a near linear scaling,][]{2007ApJ...668L..51P} but agreeing with results of global stratified simulations of \citet{2010ApJ...712.1241S}, we observe a shallow power-law relation between the transport coefficients and the net-vertical flux. If one excludes the strong-field regime where vigorous overturning motions disrupt the vertical disk structure, the dynamo coefficients are found to be surprisingly insensitive to the amount of net-vertical field.

The third subject of our paper concerns non-locality in the transport coefficients. In a seminal paper, \citet{2002GApFD..96..319B} first pointed out the possibility of scale-dependent $\alpha$ and $\eta$ tensors, and obtained the first results for stratified MRI turbulence. Even though their approach was partly hampered by very noisy data, their derived scale-dependent dynamo model provided a better fit to the simulations in terms of the cycle period. The aspect of scale-dependence -- or alternatively, of the non-locality in the relation between the electromotive forces and the mean magnetic field -- can naturally be expressed within the test-field (TF) framework, where a wavenumber is typically assigned to the TF inhomogeneity. Varying this wavenumber, \citet*{2008A&A...482..739B} first introduced a scale-dependent version of the TF method. The authors have furthermore obtained the shape of the convolution kernel that expresses the non-locality of the $\alpha$ and $\eta$ effects for the cases of simple prescribed flows, and for forced isotropic turbulence. We here apply this approach to the case of stratified MRI simulations and obtain, for the first time simultaneously, the scale-dependence of the mean-field coefficients along with their vertical profile. The knowledge of these functional dependencies will enable new, quantitative mean-field models.

While we here restrict ourselves to an isothermal equation of state, it should be remarked that allowing for redistribution of heat via strong convective motions \citep{2012ApJ...761..116B, 2013ApJ...771L..23B}, the efficiency of the dynamo can be drastically affected \citep{2013ApJ...770..100G}, potentially explaining the enhanced turbulent transport related to transient accretion outbursts \citep{2014ApJ...787....1H} and variability in the inner regions of protoplanetary disks \citep{2015MNRAS.448.3105H}. In view of devising a comprehensive mean-field model of large-scale field evolution, this has the dramatic repercussion that such a model would be required to encapsulate the disk thermodynamics. 

As a final introductory remark, we note that as in previous papers \citep[see, e.g.,][]{2015MNRAS.446.2102N}, we focus our attention to the Rayleigh stable regime, $q<2$. Our simulations beyond this value showed exponential growth of $m=0$ modes. This is expected from linear theory \citep{2012MNRAS.423L..50B,2015MNRAS.448.3697S}, and the loss of the local character of the instability clearly precludes the use of the shearing box.

This paper is organized as follows. Section~\ref{sec:methods} describes the numerical method and computational set-up and introduces the test-field concept. In Section~\ref{sec:q-dep} we discuss the shear-rate dependence, and in Section~\ref{sec:cycle} we demonstrate how this affects the cycle period of the dynamo. Section~\ref{sec:nvf-dep} deals with the NVF-dependence of the turbulent stress and the closure coefficients, and in Section~\ref{sec:non-local} we derive the spectral dependence of the mean-field effects. We discuss our results in Section~\ref{sec:summary} and draw conclusions in Section~\ref{sec:concl}.


\section{Methods}
\label{sec:methods}

The numerical simulations presented in this paper consist of solving the single-fluid MHD equations in the framework of the isothermally stratified shearing box \citep{2010MNRAS.405...41G}. Implementation details regarding the shearing-periodic boundaries, and the momentum source terms in the \NIII code are described in detail in \citet{2007CoPhC.176..652G}.

\subsection{Numerical method} 

All computations are carried out using the second-order accurate in space, and third-order accurate in time \NIII code \citep{2004JCoPh.196..393Z}, which we have furthermore extended with a Harten-Lax-van Leer-Discontinuity (HLLD) Riemann solver \citep{2005JCoPh.208..315M} that uses up-winded electromotive forces \citep{2005JCoPh.205..509G} that have been interpolated from cell faces to edges. The use of the more accurate HLLD (compared with the standard HLL solver) is expected to improve the accuracy for low Mach number flows. In the formulation with conserved variables $\rho$, $\rho\V$, and written in local Cartesian coordinates ($\xx,\,\yy,\,\zz$) the equations of ideal, isothermal MHD are:
\begin{eqnarray}
  \partial_t\rho +\nabla\cdt(\rho \V) & = & 0          \,,\\[2pt]
  \partial_t(\rho\V) +\nabla\cdt
          [\rho\mathbf{vv}+p^{\star}-\mathbf{BB}] & = &
          -\rho\nabla\Phi - 2\rho\Omega\,( \zz\tms\V ) \,,\\[2pt]
  \partial_t \B -\nabla\tms(\V\tms\B) & = & 0  \,,
\label{eq:mhd}
\end{eqnarray}
subject to the additional constraint $\nabla\cdt\B = 0$, which is preserved to machine accuracy in our constrained-transport scheme. We furthermore introduce the total pressure $p^{\star}\equiv p+\frac{1}{2}\B^2$, and specify the effective potential 
\begin{equation}
  \Phi(x,z) = -q\Omega^2x^2 + \frac{1}{2}\Omega^2z^2
\end{equation}
in the local frame of reference, rotating with a fixed angular frequency $\OO\equiv\Omega\zz$. Note that unlike in our previous papers, we here absorb the minus sign into the definition of the shear-parameter, $\ q\equiv -{\rm d}\ln \Omega/{\rm d}\ln r$.

For all simulations presented here, we use the same box size of $L_x\times L_y\times L_z = H\times\pi H\times 6H$ with a fixed linear resolution of $\sim32/H$ in all three space dimensions. This amounts to $32\times 96\times 192$ cells in the radial, azimuthal, and vertical coordinate directions, respectively. The smaller box size compared to, for instance, the recent study of \citet{2015MNRAS.446.2102N} -- who used a box with $4H\times 4H\times 8H$ at a resolution of $24/H$ -- is chosen to accommodate for a larger number of simulations, and a longer simulation time of typically 800 orbits (with period $P\equiv 2\pi\Omega^{-1}$). This allows us to obtain much better statistics, which is especially relevant for the dynamo coefficients obtained with the test-field method. We have furthermore performed a reference simulation with twice the resolution, $64/H$, which we ran for $\sim 270$ orbits. The studied quantities are typically converged at the 20--40\% level (also cf. Tab.~\ref{tab:results}).

We note that we consciously neglect explicit viscous or resistive terms \citep[but see discussion in][for unstratified systems]{2007A&A...476.1123F} and rely on the truncation error of our numerical scheme to approximate the effects of small-scale energy dissipation. This implies that our approach is better suited for addressing the dynamics of accretion disks that do not have extreme ratios of viscous to resistive microscopic dissipation coefficients \citep{2008ApJ...674..408B}. The dependence of the MRI on dimensionless numbers is an active field of research \citep[e.g.][]{2007A&A...476.1123F,2007MNRAS.378.1471L,2008ApJ...684..498P,2010A&A...516A..51L}. At low to intermediate Reynolds numbers, and already in the absence of vertical stratification, the MRI furthermore exhibits interesting dynamical behavior \citep{2011PhRvE..84c6321H,2013JFM...731....1R,2015A&A...575A..14R}.

Even though the numerical timestep may in practice be restricted by the Alfv{\'e}n speed high up in the disk corona (especially in models with a net-vertical field), we nevertheless apply an orbital advection scheme \citep[cf.][]{2010ApJS..189..142S}. This has the advantage that the truncation error of our Godunov scheme, which sensitively depends on the advection velocity, remains uniform throughout the simulation domain. For the interpolation of non-integer displacements, we use the Fourier method suggested by \citet{2009ApJ...697.1269J}. Rather than using a so-called Alfv{\'e}n limiter \citep[see, e.g.][]{2000apj...534..398m}, we instead include an artificial mass diffusion term \citep[described in detail in][]{2011MNRAS.415.3291G} to circumvent undue time-step constraints resulting from low-density regions in the upper disk corona.

\subsection{Initial conditions} 

We initialize the fluid velocity with the background equilibrium solution $\V_0=q\Omega\,x\,\yy$, which -- in the orbital advection scheme -- amounts to setting the perturbed velocity identically to zero. Neglecting the pressure support of the weak magnetic field, we obtain the initial density profile by solving for hydrostatic equilibrium in the vertical direction, yielding
\begin{equation}
  \rho(z) = \rho_0\, {\rm exp}\,\left(-\frac{z^2}{2H^2}\right)\,.
  \label{eq:hydrostat}
\end{equation}
This equation also serves to define our adopted convention for the pressure scale-height, $H$. We start our simulation by seeding the density (and hence the pressure) with a standard white-noise perturbation of $1\%$ rms amplitude.

The initial magnetic field in all the simulations has the form
\begin{equation}
  \B=\left[\, B_0^{\rm ZNF}\,
    \sin\left(\frac{2\pi x}{L_x}\right)\,\sqrt{\frac{p(z)}{p_0}}
    + B_0^{\rm NVF}\,\right]\,\zz\;+B_x(x,z)\,\xx\,,
\end{equation}
which is designed to obtain a transition into the turbulent state that is as uniform as possible. By scaling the vertical field with a factor $(p(z)/p_0)^{1/2}$, we obtain $\mn{\beta}_{\rm P}={\rm const}$ (which we set to a value of $1600$, initially), implying that the wavelength of the most unstable MRI mode, $\lambda_{\rm MRI}$, becomes independent of height. Mandated by the divergence-free nature of the magnetic field, and because $B_z=B_z(x,z)$, we have to add a corresponding radial field, $B_x(x,z)$. In practice, this is done by specifying a suitable vector potential $\bm{A}\equiv A_y(x,z)\,\yy$. In the cases where we are interested in the shear-rate dependence of the resulting turbulence properties, the initial magnetic configuration is of the zero-net-flux (ZNF) type, i.e., $B_0^{\rm NVF}=0$. In simulations with a net-vertical field (NVF), the amplitude of the component $B_0\equiv B_0^{\rm NVF}$ is typically stated in terms of the midplane value, $\betap^{\rm mid}\equiv 2p_0/B_0^2$, of the plasma parameter, $\betap$.

\subsection{Boundary conditions} 

Reflecting the local character of the box geometry, the horizontal boundary conditions (BCs) are of the usual sheared-periodic type. As in previous work, we modify the hydrodynamic fluxes (and the azimuthal component of the electromotive force) to retain the conservation properties of the finite-volume constrained transport scheme \citep{2007CoPhC.176..652G}. In the vertical direction, we adopt stress-free BCs, that is, $\partial_z v_x=\partial_z v_y=0$. We compute the density values of adjacent grid cells in the $z$~direction to be in hydrostatic equilibrium (neglecting, however, any contribution from the magnetic pressure). We furthermore allow outflow of gas through the $z$~boundaries, while preventing in-fall of material from outside the domain. In order to counter-act the severe mass loss owing to the emerging disk wind (even in the ZNF case), we continuously rescale the mass density to re-instate the original mass, $M_0$, within the box. This is done in the same way as in earlier work \citep{2012MNRAS.422.1140G}, which also adopted an isothermal equation of state. We remark that, in situations with a strong NVF, the mass loss per orbit can approach 100\% of the mass contained within the box. Nevertheless, such a replenishing of material can be thought of as a natural consequence of radial mass transport within a global disk if the system is in a quasi-stationary state on average. For the magnetic field, we use the standard normal-field $\partial_z B=0,\,B_x=B_y=0$ (sometimes referred to as ``pseudo-vacuum'') boundary condition in the vertical direction. Forcing the horizontal field to zero at the vertical boundaries has the effect of creating an additional pressure gradient, potentially helping the already existing outflow to become super-Alfv{\'e}nic.

\subsection{The test-field method} 
\label{sec:tf_method}

One central objective of this paper is to study how the dynamo closure coefficients of the mean-field dynamo\footnote{These are thought to be responsible for the evolution of the large-scale magnetic fields via a mean-field dynamo equation -- see, for instance, \citet*{2012SSRv..169..123B} for a topical review.} depend on various input parameter such as the shear-rate and the strength of the applied net-vertical field. As in previous work, we heavily rely on the now well-established test-field method \citep{2005AN....326..245S,2007GApFD.101...81S} to infer closure coefficients. These describe, for instance, the classical $\alpha$~effect \citep{1955ApJ...122..293P}, turbulent (``diamagnetic'') pumping, $\gamma$, and enhanced resistivity, $\eta_{\rm T}$, which have long been recognized to affect the long-term evolution of magnetic fields in turbulent systems \citep{1980opp..bookR....K}. The TF method provides us with time-steady measurements of the $\alpha$ and $\eta$ tensors as functions of the height in the disk.\footnote{Further details are documented in section~3.1 of \citet{2008AN....329..619G}.} Obtaining these coefficients, for instance, as a function of the shear-rate -- while building on existing theory of mean-field dynamos -- allows us to make meaningful predictions about whether the characteristic butterfly diagram can indeed be described by a classical $\alpha\Omega$~dynamo, for which a dispersion relation can be derived \citep[e.g.,][]{2005PhR...417....1B}.

The method, as it is applied here, can be classified as ``quasi-kinematic'' \citep{2008ApJ...687L..49B} in the sense that it only solves for magnetic fluctuations $\B'(\bm{x},t)$. This is done via an additional set of induction equations solved simultaneously with the main simulation. The measurement procedure is in fact similar to the approach used by \citet{2009A&A...504..309L}, who impose a sinusoidal magnetic field perturbation and infer the correlation between the resulting electromotive force and the current. The difference in the TF method is that the perturbation is \emph{virtual} in the sense that the simulation remains unaware of the applied field. While the additional equations properly take into account the actual turbulent velocity $\U'(\bm{x},t)$ of the main simulation, they ignore contributions in the \emph{mean} electromotive force, $\EMF \equiv \overline{\U'\tms \B'}$ that derive from a non-linear evolution of the kinematic fluctuations \citep[see, however,][]{2010A&A...520A..28R}. Employing simulations of forced turbulence, \citet{2008ApJ...687L..49B} have found that the quasi-kinematic method remains valid for cases that do not contain dominant magnetic fluctuations that are already present in the \emph{absence} of mean fields.

The shearing-box approximation, that we use here, eliminates characteristic gradients in the radial and azimuthal directions. We accordingly consider mean fields that depend on the vertical coordinate only. Using the Lagrangian velocity $\U \equiv \V + q\Omega x \yy$, and defining $\U=\mU+\U'$, as well as $\B=\mB+\B'$, the mean-field induction equation can be derived as
\begin{equation}
  \partial_t \mB(z) - \nabla \times \left[\ \mU(z)\tms\mB(z)
    + \EMF(z) - (q\Omega x\yy)\tms\mB(z)\ \right] = 0\,.
  \label{eq:MF_ind}
\end{equation}
This equation, via the term $\EMF(z)$, still depends on the fluctuating quantities $\U'$ and $\B'$. To solve it independently in the form of a mean-field dynamo model, one needs to parametrize $\EMF$ in terms of the mean quantities, $\mU$, and $\mB$, and statistical properties of the fluctuating quantities alone. Following \citet{2008an....329..725b}, we apply a local formulation for the EMF
\begin{equation}
  \EMF_i(z) = \alpha_{i\!j}(z)\ \mn{B}_j(z)
  \ -\ \eta_{i\!j}(z)\ \varepsilon_{\!jzl}\,\partial_z \mn{B}_l(z)\,,
  \label{eq:closure}
\end{equation}
where the properties of the turbulence are encapsulated in the $\alpha_{i\!j}(z)$ and $\eta_{i\!j}(z)$ tensors, relating the turbulent electromotive force to the \emph{local} values of the mean magnetic field, $\mn{B}(z)$, and its derivatives.

The approach embodied in equation~(\ref{eq:closure}) formally requires separation of scales, such that non-local dependencies in space, and so-called ``memory effects'' \citep{2009ApJ...706..712H} can be neglected. Accounting for such effects requires to replace the simple multiplicative relation in equation~(\ref{eq:closure}) by a convolution integral, as described in detail via equations~(\ref{eq:closure_convolutions}) -- (\ref{eq:kernel_real_space}) in Section~\ref{sec:non-local} \citep[also cf.][]{2008A&A...482..739B}. In this case, the closure coefficients formally become convolution kernels. For the case of isotropic forced turbulence, \citet{2008A&A...482..739B} have found that the kernels can be approximated by a Lorentzian in Fourier space, implying exponential decay in real space. This was also the case for helical turbulence with shear \citep{2009A&A...495....1M} and for passive scalar diffusion \citep{2010PhRvE..82a6304M} -- which may imply that this is a generic feature.

A test field is thought to be an external inhomogeneity (caused in reality by the large-scale magnetic field), in turn leading to non-vanishing correlations in $\U'\times\B'$. In the case of a horizontally-periodic, stratified box, it is common to chose the test fields to be functions of $z$, and apply averages over the $x$ and $y$ directions. Despite the non-periodic boundaries in the vertical direction, we chose harmonic functions with a specified wavenumber $k_z^{\rm TF}$. In our previous studies, we were mainly interested in large-scale mean fields, and hence restricted ourselves to a single value $k_z^{\rm TF}=k_1\equiv 2\pi/L_z$. In sections \ref{sec:q-dep} and \ref{sec:nvf-dep}, we again focus on a single value, namely $k_z^{\rm TF}=k_2$. To study a potential dependence on the scale-size of the mean magnetic field, in section \ref{sec:non-local}, we extend this approach to a range of values $k_z^{\rm TF}=k_1 \dots k_{32}$ (while keeping the NVF and shear rate fixed for this set of runs).


\begin{table*}\begin{center}
\caption{Summary of simulation results.
  \label{tab:results}}
\begin{tabular}{ccccccccccccc}\hline
    $q$                     & $\bar{B}_z^{\rm NVF}$
  & $T_{R\phi}^{\rm Reyn}$  & $T_{R\phi}^{\rm Maxw}$  & $T_{R\phi}^{\rm ratio}$
  & $P_{\rm cyc}$           & $\tauc$                 & $u_{\rm rms}$
  & $\alpha_{yy}^{\rm peak}$& $\alpha_{yy}^{\rm buoy}$
  & $\eta_{yy}^{\rm peak}$  & $\eta_{yy}^{\rm mid}$ 
  \\[4pt]
                            & $[\beta_{\rm p\,800}]$
  & $[10^{-3}p_0]$          & $[10^{-3}p_0]$          & 
  & $\![2\pi\Omega^{-1}]\!$ & $[\Omega^{-1}]$         & $[H\Omega]$
  & $\![10^{-2}H\Omega]\!$  & $\![10^{-2}H\Omega]\!$ 
  & $\![10^{-2}H^2\Omega]\!$& $\![10^{-2}H^2\Omega]\!$ 
  \\[1pt]
  \hline\\[-6pt]
    0.3                    & $0.06$
  & $ 0.01\pm 0.00$        & $ 0.24\pm 0.10$        & $ 36.7$
  & $ 48.1$                & $ 0.147$               & $ 0.10\pm 0.03$
  & $ 0.106$               & $-0.002$
  & $ 0.091$               & $ 0.019$\\
    0.6                    & $0.06$
  & $ 0.07\pm 0.03$        & $ 1.13\pm 0.37$        & $ 16.2$
  & $ 25.2$                & $ 0.091$               & $ 0.26\pm 0.08$
  & $ 0.386$               & $-0.011$
  & $ 0.581$               & $ 0.090$\\
    0.9                    & $0.06$
  & $ 0.23\pm 0.09$        & $ 2.12\pm 0.70$        & $  9.3$
  & $ 12.6$                & $ 0.067$               & $ 0.41\pm 0.12$
  & $ 0.673$               & $-0.036$
  & $ 0.972$               & $ 0.155$\\
    1.2                    & $0.06$
  & $ 0.76\pm 0.26$        & $ 4.47\pm 1.35$        & $  5.9$
  & $  9.5$                & $ 0.046$               & $ 0.68\pm 0.15$
  & $ 0.936$               & $-0.059$
  & $ 1.478$               & $ 0.314$\\
    1.5                    & $0.00$
  & $ 1.49\pm 0.46$        & $ 5.88\pm 1.63$        & $  3.9$
  & $  7.5$                & $ 0.039$               & $ 0.86\pm 0.14$
  & $ 1.048$               & $-0.014$
  & $ 1.530$               & $ 0.556$\\
    1.8                    & $0.06$
  & $ 4.16\pm 1.02$        & $11.75\pm 2.88$        & $  2.8$
  & $  5.9$                & $ 0.027$               & $ 1.31\pm 0.06$
  & $ 1.200$               & --
  & $ 1.513$               & $ 1.232$\\[4pt]
    1.5$^{\,\star}$\hspace{-5.5pt} & $0.00$
  & $ 1.14\pm 0.37$        & $ 4.99\pm 1.70$        & $  4.4$
  & $  6.8$                & $ 0.016$               & $ 0.84\pm 0.01$
  & $ 0.867$               & $-0.015$
  & $ 1.233$               & $ 0.337$\\[4pt]
    1.5                    & $0.01$
  & $ 1.49\pm 0.44$        & $ 5.88\pm 1.55$        & $  3.9$
  & $  7.3$                & $ 0.039$               & $ 0.69\pm 0.03$
  & $ 1.031$               & $-0.017$
  & $ 1.541$               & $ 0.533$\\
    1.5                    & $0.02$
  & $ 1.38\pm 0.39$        & $ 5.48\pm 1.35$        & $  4.0$
  & $  7.1$                & $ 0.038$               & $ 0.67\pm 0.03$
  & $ 1.107$               & $-0.020$
  & $ 1.587$               & $ 0.494$\\
    1.5                    & $0.04$
  & $ 1.58\pm 0.55$        & $ 6.26\pm 2.01$        & $  4.0$
  & $  7.2$                & $ 0.036$               & $ 0.72\pm 0.02$
  & $ 1.023$               & $-0.003$
  & $ 1.544$               & $ 0.586$\\
    1.5                    & $0.08$
  & $ 1.70\pm 0.53$        & $ 6.75\pm 1.99$        & $  4.0$
  & $  7.1$                & $ 0.035$               & $ 0.76\pm 0.02$
  & $ 1.100$               & $-0.014$
  & $ 1.614$               & $ 0.603$\\[4pt]
    1.5                    & $0.16$
  & $ 2.20\pm 0.57$        & $ 8.81\pm 2.00$        & $  4.0$
  & $  6.4$                & $ 0.030$               & $ 0.90\pm 0.03$
  & $ 1.149$               & --
  & $ 1.833$               & $ 0.703$\\
    1.5                    & $0.32$
  & $ 3.57\pm 0.96$        & $12.78\pm 2.73$        & $  3.6$
  & $  5.9$                & $ 0.018$               & $ 1.20\pm 0.01$
  & $ 0.950$               & --
  & $ 2.475$               & $ 0.682$\\
    1.5                    & $0.64$
  & $ 5.59\pm 3.00$        & $15.06\pm 6.17$        & $  2.7$
  & --                     & $ 0.012$               & $ 1.32\pm 0.03$
  & $ 0.655$               & --
  & $ 4.425$               & $ 0.436$\\
    1.5                    & $1.28$
  & $ 5.32\pm 9.49$        & $12.74\pm11.33$        & $  2.4$
  & --                     & $ 0.019$               & $ 1.11\pm 0.04$
  & $ 0.604$               & --
  & $ 4.231$               & $ 0.550$\\
 \hline
\end{tabular}
\end{center}
  \parbox[t]{2.05\columnwidth}{\footnotesize \emph{Notes:} Here, the shear rate is defined as $q\equiv -{\rm d}\ln\Omega/{\rm d}\ln r$, that is, $q=1.5$ for Keplerian rotation. The net-vertical field, $\bar{B}_z$, is given in units that are multiples of the field strength resulting in a midplane plasma parameter, $\betap^{\rm mid}=800$. The Reynolds and Maxwell stresses are computed as correlations of fluctuating quantities. The cycle period, $P_{\rm cyc}$, of the dynamo butterfly diagram is obtained as described in the text. The correlation time, $\tauc$, refers to the classic estimate of the turbulent diffusivity (cf. Fig.~\ref{fig:q_tau_c}). In the chosen units, $u_{\rm rms}$ is equivalent of a turbulent Mach number. Dynamo coefficients labeled `peak' (`buoy') correspond to open squares (triangles) in Fig.~\ref{fig:q_dynamo_yy}. The same holds for the turbulent diffusivity, which is plotted in Fig.~\ref{fig:q_eta_yy}, accordingly. The run marked with the asterisk ($^\star$) is a double-resolution reference run with $64/H$ grids, indicating that results are generally converged at the 20--40\% level.}
\end{table*}


\section{Results}
\label{sec:results}

Our simulation results are summarized in Table~\ref{tab:results}, where the first two columns show the input parameters, and the remaining columns list characteristic quantities, and representative mean-field dynamo coefficients. We start by discussing the effect of varying the shear-parameter, $q$, while keeping the net-vertical field fixed at a low value, corresponding to a midplane $\betap\simeq 2\times 10^{5}$. Conversely, we then study the dependence on the NVF, when keeping the shear rate fixed at its Keplerian value, $q=1.5$. Comparing the results for the turbulent transport coefficients with the inferred mean-field coefficients promises to provide insights into how the two phenomena are potentially linked. Lastly, we study potential non-local contributions to the mean-field dynamo tensors, keeping both $q$ and $\betap$ fixed.

\subsection{Dependence on the shear-rate} 
\label{sec:q-dep}

One criticism of the classic ``enhanced'' viscosity picture has been that the stress associated with eddy viscosity is assumed to depend \emph{linearly} on the shear rate \citep[see discussion in][]{2008MNRAS.383..683P}.  This has been shown to be an invalid assumption, since the efficiency of the turbulent angular-momentum transport depends on the shear rate in a more subtle way, as has also been found by many authors \citep{1996MNRAS.281L..21A,2001A&A...378..668Z,2009AN....330...92L,2015MNRAS.446.2102N}.  The upper panel of Fig.~\ref{fig:q_stress} illustrates the dependence of the turbulent stresses on the shear parameter $q$. Our results are in excellent agreement with those of \citet{2015MNRAS.446.2102N}, who recently studied the ZNF case, and also confirm the scaling of the total stress with the shear-to-vorticity ratio $q/(2-q)$ that had first been suggested by \citet{1996MNRAS.281L..21A} based on their numerical findings. The scaling of the hydrodynamic Reynolds stress is somewhat steeper, leading to the known $(4-q)/q$ scaling for the ratio of the magnetic-to-hydrodynamic stresses \citep{2006MNRAS.372..183P}, which we plot in the bottom panel of Fig.~\ref{fig:q_stress}. Our results are consistent with the stratified simulations of \citet{2001A&A...378..668Z} and \citet{2015MNRAS.446.2102N}, and the unstratified simulations of \citet{2008MNRAS.383..683P} and \citet{2009AN....330...92L}. Quite remarkably, this ``signature'' can be traced back to the eigenmode structure of the linear MRI problem in the limit of ideal MHD \citep{2006MNRAS.372..183P}.

\begin{figure}
  \center\includegraphics[width=0.95\columnwidth]{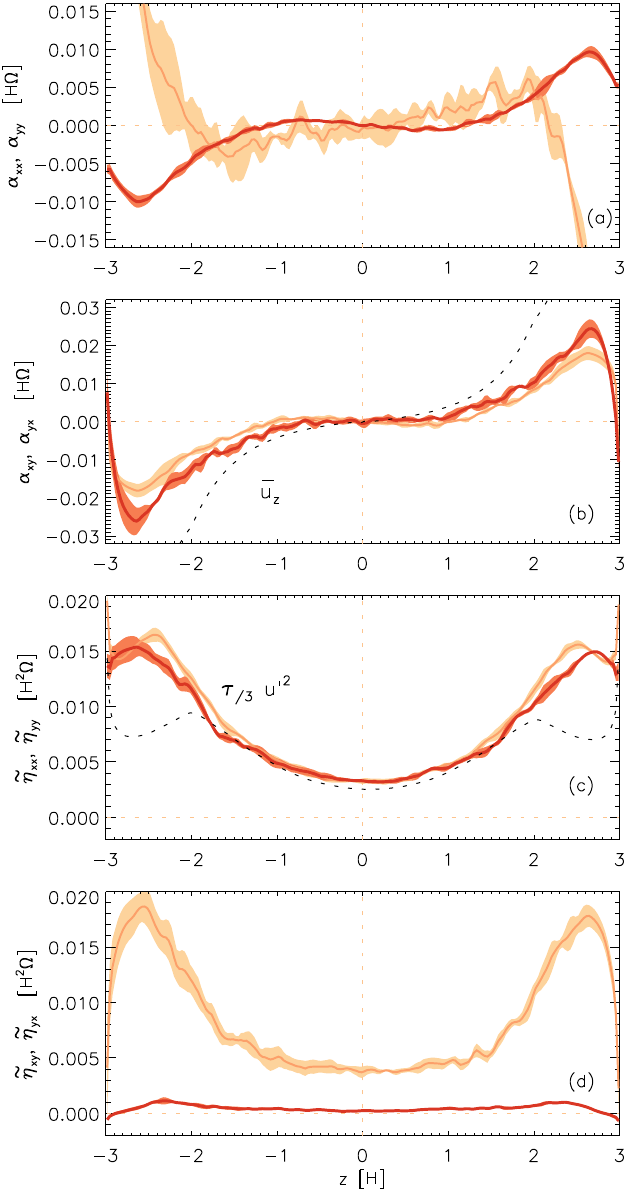}
  \caption{Turbulent mean-field coefficients based on the TF analysis ($t=100-800$ orbits). As a representative example, we show the curves for the model with $q=1.2$. The labels on the vertical axes indicate the quantities plotted in light ($\alpha_{xx}$, $\alpha_{xy}$, $\eta_{xx}$, $\eta_{xy}$) or dark ($\alpha_{yy},\dots$) shades. The mean vertical flow velocity, $\mn{u}_z$, in panel~(b), and the quasi-linear estimate for the isotropic turbulent diffusion, in panel~(c), are over-plotted as dashed lines.}
  \label{fig:dyn_q12}
\end{figure}

In order to establish whether the saturated amplitude of the turbulent transport coefficients depends on the presence of a mean-field dynamo, we have measured the coefficients that characterize such activity.  As in previous work \citep{2005an....326..787b,2008an....329..725b,2010MNRAS.405...41G,2013ApJ...770..100G}, we assume a simple local closure between the turbulent electromotive force, $\EMF$, and the mean magnetic field, $\mB$ (see section \ref{sec:tf_method} for details), and use the test-field method to obtain vertical profiles of various tensor components. 

The results obtained for the dynamo coefficients with the TF method are illustrated in Fig.~\ref{fig:dyn_q12} for the model with $q=1.2$, and $\bar{B}_z=0.06$. The first panel, (a), shows the diagonal elements of the dynamo $\alpha$~tensor. For homogeneous isotropic turbulence, this $\alpha$~effect can be shown to be proportional to the negative kinetic helicity \citep{1980opp..bookR....K}. For kinetically driven, stratified, rotating turbulence, analytic theory predicts that $\alpha$ will emerge naturally from the inhomogeneity introduced by the stratification combined with the anisotropy caused by the rotation. The magnitude of the effect is predicted to be proportional to the gradient in the mean density combined with some power of the gradient in the turbulent velocity \citep[see the discussion in][]{2013ApJ...762..127B}.

Because of the dominant shear, $\alpha_{xx}$ is expected to play a minor role, since azimuthal field can efficiently be generated from radial field by the background shear flow -- the so-called $\alpha\Omega$ dynamo regime. Then, the $\alpha_{yy}$ element plays an important role in producing radial fields from azimuthal fields via the vertical variation of $\EMF_y=\alpha_{yy}\,\bar{B}_y$. Note that $\alpha_{yy}$ shows opposite slopes near the disk midplane and in the disk corona.

Based on the known behavior of $\alpha\Omega$ dynamos, \citet{1998tbha.conf...61B} has argued that a negative slope near the midplane is required to explain the upward migration observed in the characteristic ``butterfly'' diagram in stratified MRI simulations \citep[also see discussion in][]{2010MNRAS.405...41G}. The results presented here challenge this hypothesis, since we find the negative slope to vanish for $q\simgt 1.5$, whereas the dynamo pattern remains unaffected (see discussion in Sect.~\ref{sec:cycle}). This finding potentially implies that another effect may be responsible for the observed pattern speed (and direction of propagation).

A peculiarity of the MRI dynamo coefficients that had been noted before \citep{2008an....329..725b} can be seen in the next panel of Fig.~\ref{fig:dyn_q12} -- panel (b) shows the off-diagonal elements of the $\alpha$~tensor, which are found to be \emph{symmetric} (i.e., $\alpha_{xy}\simeq\alpha_{yx}$). In isotropic kinematically-forced turbulence, one typically expects \citep[e.g.,][]{1980opp..bookR....K} that the dominant contribution to the off-diagonal elements are given by so-called diamagnetic pumping, $\gamma = -\nabla\eta_{\rm T}/2$, reflected in \emph{anti-}symmetric tensor elements, $\alpha_{yx}=-\alpha_{xy}$.  Such an effect implies an expulsion of magnetic flux from strongly turbulent regions into more quiescent regions, and has for instance been observed in kinetically forced interstellar turbulence \citep{2008A&A...486L..35G}. The fact that both coefficients show the same sign implies that this effect is significantly anisotropic in MRI turbulence, that is to say that $B_x$ and $B_y$ are affected differently.

The diagonal elements of the diffusivity tensor are shown in panel (c) of Fig.~\ref{fig:dyn_q12}. Complementary to the result of \citet{2009A&A...504..309L}, who find the diffusion tensor to be anisotropic when comparing the radial diffusion of the azimuthal and vertical field components, the coefficients measured from our simulations approximately satisfy $\eta_{xx}\simeq \eta_{yy}$ -- these coefficients however describe \emph{vertical} diffusion of the azimuthal and radial field components, respectively. Isotropy is not necessarily expected for a shear flow with a dominant azimuthal field component \citep{1993ApJ...404..773V}, and the deviation is indeed most pronounced in the more strongly magnetized disk corona ($z\simgt 2\,H$). As previously found \citep{2009A&A...504..309L,2010MNRAS.405...41G}, the $\eta$~tensor, moreover, has significant off-diagonal contributions, shown in panel (d). The $\eta_{yx}$ term has the potential to couple radial and azimuthal fields (via their vertical gradients), and hence may be responsible for pattern propagation and/or cyclic behavior \citep{2010MNRAS.405...41G,2011PhRvE..84c6321H}. The three dominant coefficients of the diffusivity tensor (i.e., $\eta_{xx}$, $\eta_{yy}$ and $\eta_{xy}$) closely follow the theoretical expectation
\begin{equation}
  \eta \simeq \frac{1}{3}\,\tauc\,u_{\rm rms}^2
  \label{eq:eta}
\end{equation}
within the disk body ($z\simlt 2\,H$, see panel (c) in Fig.~\ref{fig:dyn_q12}), where we have assumed a constant correlation time, $\tauc$, as well as isotropic velocity perturbations. In the low-$\beta$ disk corona, magnetic fluctuations are likely dominant, perhaps explaining the deviation from the kinematic result. We will later use equation~(\ref{eq:eta}) to obtain an independent estimate of the turbulent coherence time $\tauc$ as a function of shear rate (see Fig.~\ref{fig:q_tau_c}, below).

\begin{figure}
  \center\includegraphics[width=\columnwidth]{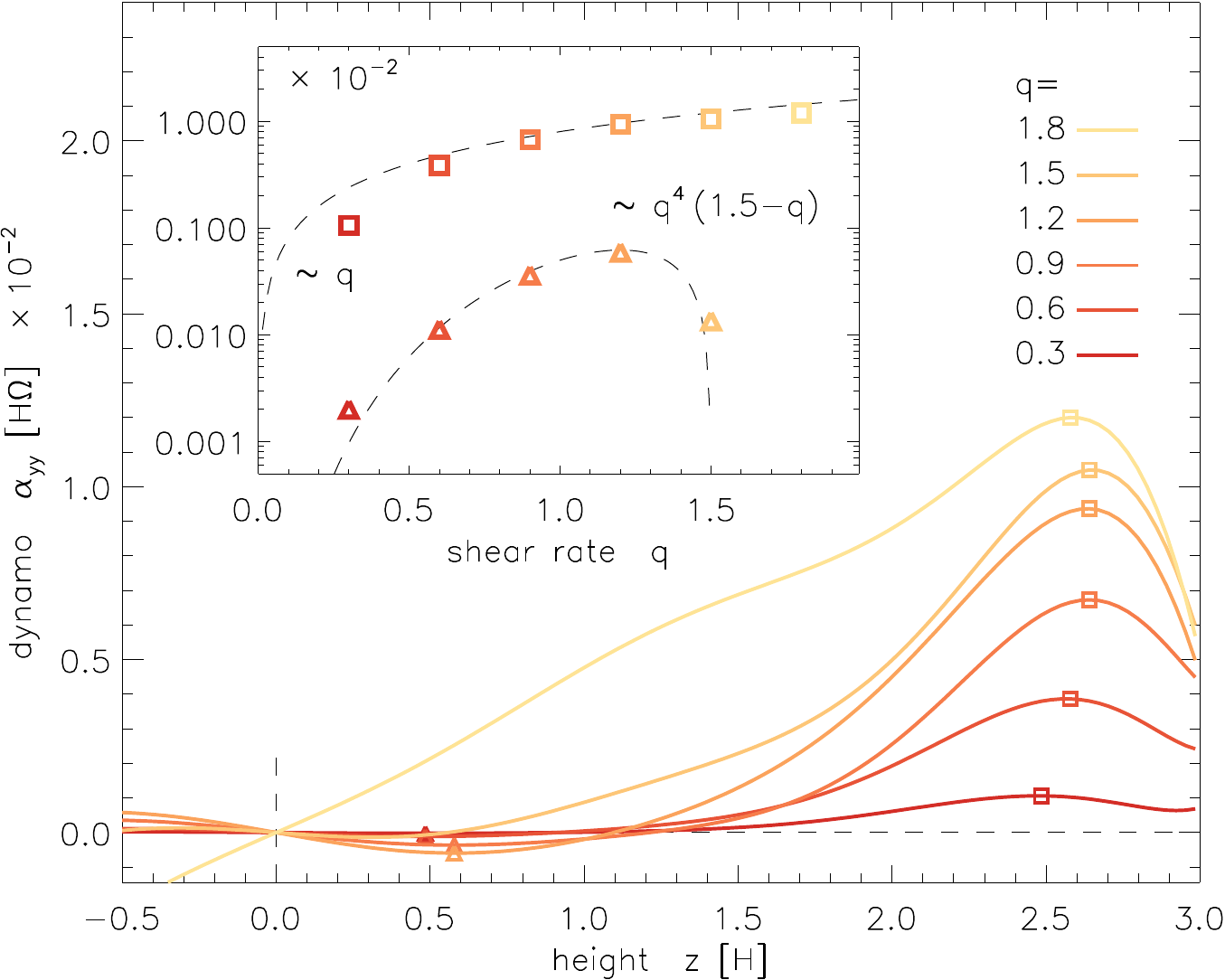}
  \caption{Vertical profiles (main plot), and shear-dependence (inset) of the $\alpha_{yy}(z)$ coefficient. Amplitudes for the kinematic dynamo (squares), and the buoyancy-driven dynamo (triangles) have been inferred from peak values as indicated. Heuristic approximations for the $q$-dependence are also shown (dashed lines). When approaching the Rayleigh limit ($q\!\rightarrow\!2$), field-line buoyancy appears to be more and more dominated by hydrodynamic effects.}
  \label{fig:q_dynamo_yy}
\end{figure}

In Fig.~\ref{fig:q_dynamo_yy}, we compare the vertical profiles of the $\alpha_{yy}$ coefficient for the six simulation runs with varying shear rate.\footnote{To facilitate a direct comparison, we have filtered the profiles by expanding them into Legendre polynomials up to order 12, and set the even (odd) components to zero for $\alpha$ ($\eta$) to enforce the analytically expected symmetry.}  For lack of a better alternative, we resort to describing the strength of the effect simply by peak amplitudes (squares). The dependence of these amplitudes is plotted in the inset of Fig.~\ref{fig:q_dynamo_yy} (again using open squares), where we see that the dominant positive $\alpha$~effect scales approximately linearly with the shear rate. The shear-rate dependence is markedly different for the weak \emph{negative} $\alpha$~effect near the disk midplane (triangles).  This effect is thought to originate from correlated magnetic fluctuations \citep{1998tbha.conf...61B} that are associated with the buoyant rise of regions with strong azimuthal field. Based on quasi-linear theory, \citet{2000A&A...362..756R} had predicted that this effect should only be present for a finite range of the shear parameter -- given by the condition $0.75 < q < 1$ in their analysis; cf. their equations~(14) and (20). While the precise parameter range is different from this analysis by a small factor, we indeed confirm this predicted behavior. We empirically describe the $q$~dependence to be proportional to $q^4\,(1.5-q)$, and we speculate that the discrepancy may be related to the assumed spectrum of the magnetic fluctuations in their derivation; cf. \citet{2000A&A...362..756R}, equations~(17) and (18).

The result illustrated in Fig.~\ref{fig:q_dynamo_yy} qualitatively agrees with that of \citet{2001A&A...378..668Z}, who also found $\alpha$ to vanish at a finite value of $q$, and whose values scale similarly to ours -- albeit their effect is stronger by about a factor of four (cf. their table~1). In general terms, the existence of a finite interval for this effect derives from the individual contributions due to the shear and Coriolis term entering with opposite signs \citep{2000A&A...362..756R}. We note that this is however qualitatively different for the positive $\alpha$~effect further up in the disk \citep[which may be more akin to the mechanism derived in][]{1993A&A...269..581R}.

\begin{figure}
  \center\includegraphics[width=\columnwidth]{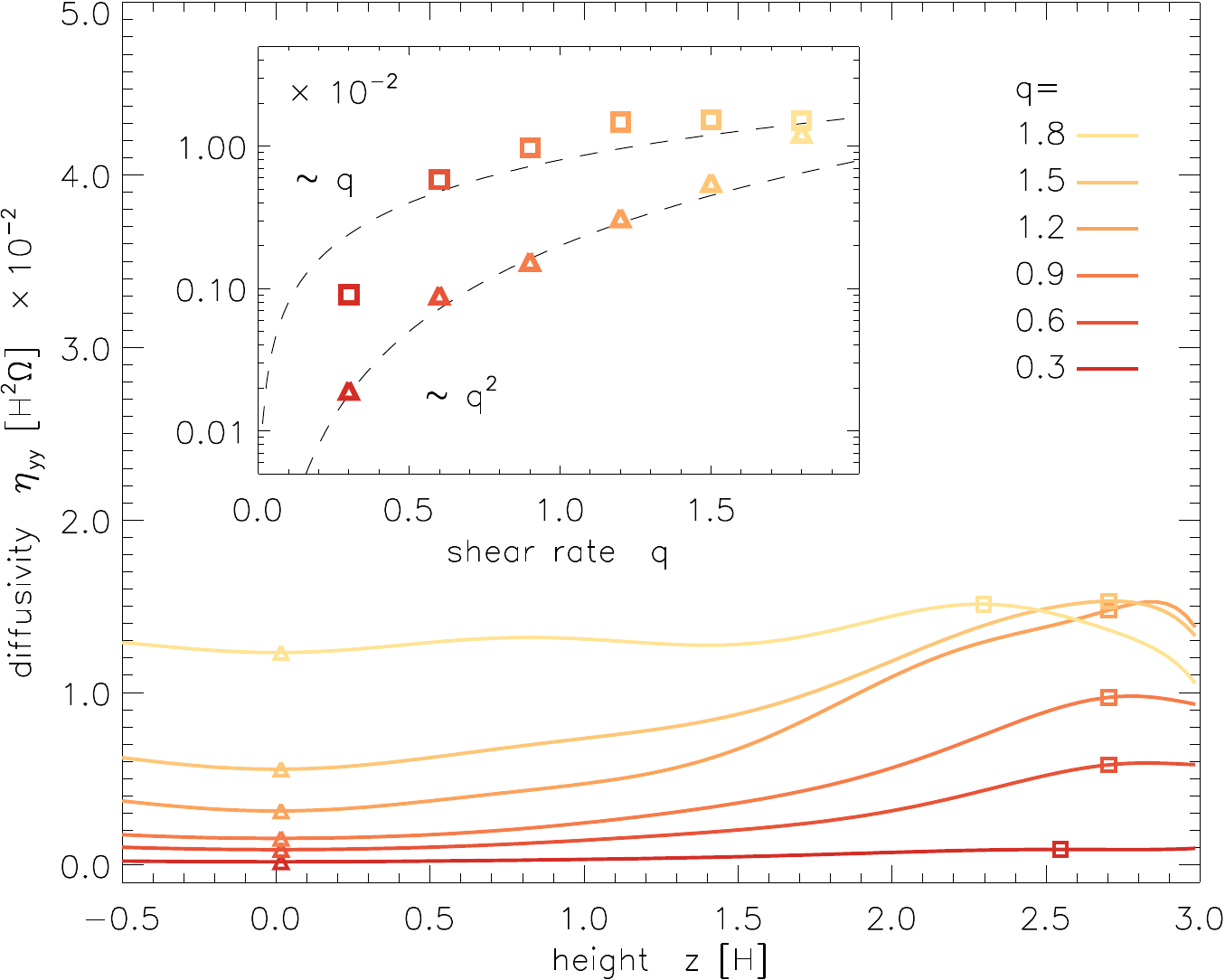}
  \caption{Same as Fig.~\ref{fig:q_dynamo_yy}, but for the diffusivity coefficient $\eta_{yy}(z)$. Peak values in the disk corona (squares) roughly scale linear with $q$, while the midplane values show a deviant scaling with $q^2$, instead. Qualitatively similar trends are found in the other three tensor coefficients.}
  \label{fig:q_eta_yy}
\end{figure}

The vertical profiles and amplitudes of the turbulent diffusivity $\eta_{yy}(z)$ are shown in Fig.~\ref{fig:q_eta_yy}. We again distinguish between peak values in the upper disk layers (squares) and values near the midplane (triangles). The former effect scales roughly with the shear rate, $q$, whereas the latter scales with $q^2$, instead.
\begin{figure}
  \center\includegraphics[width=0.95\columnwidth]{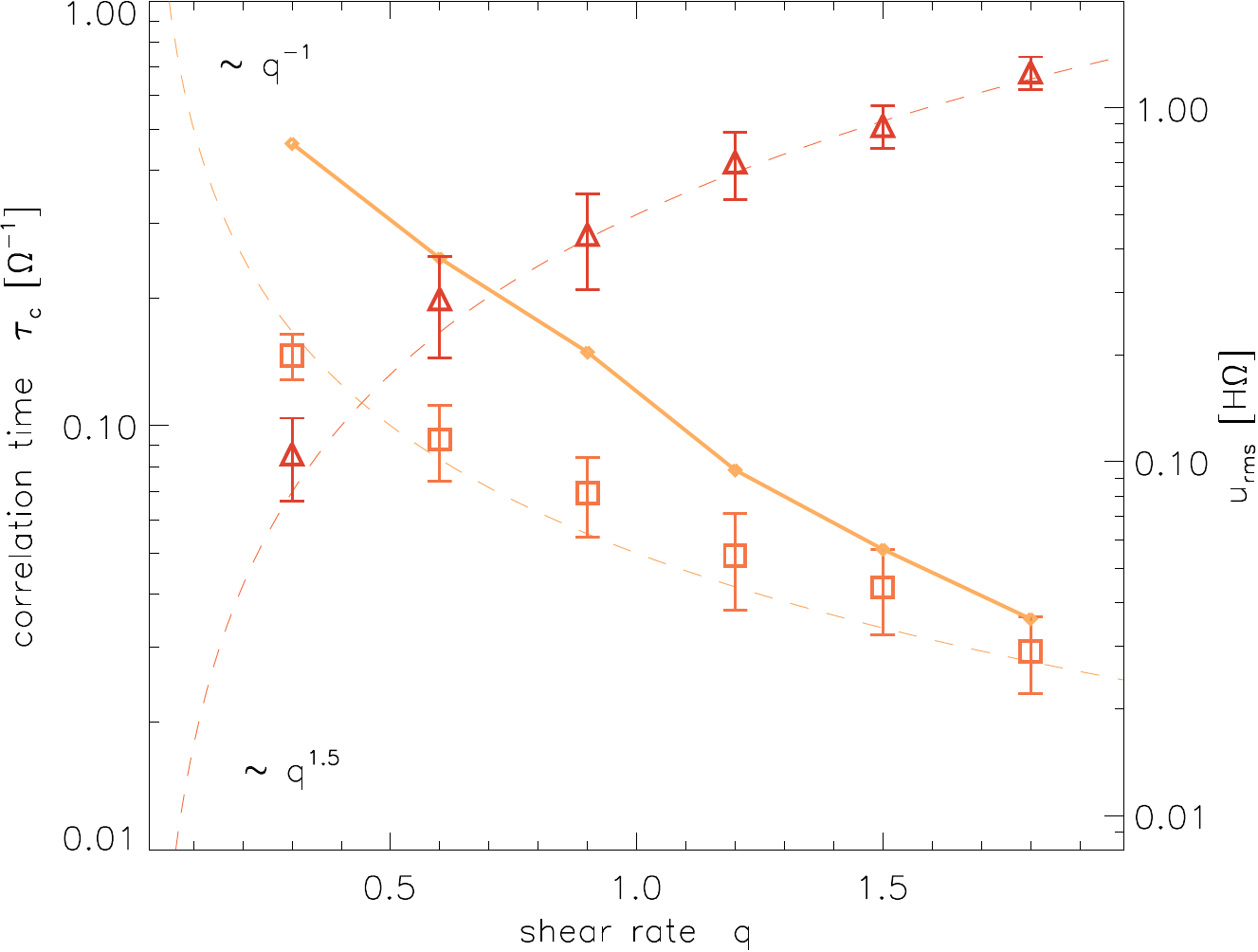}
  \caption{Shear-rate dependence of the correlation time $\tauc$ (squares) derived from the diffusivity via assuming $\eta_{yy}\simeq \tauc/3\,u_{\rm rms}^2$. The solid line shows an attempt to independently obtain $\tauc$ from the auto-correlation function of kinetic helicity. For comparison, we also show the scaling of $u_{\rm rms}$ itself (triangles), which scales approximately with $q^{3/2}$, explaining the quadratic scaling of $\eta_{yy}\,(z=0)$ seen in Fig.~\ref{fig:q_eta_yy} above.}
  \label{fig:q_tau_c}
\end{figure}
This scaling can be understood in terms of the scaling of $\tauc$, and $u_{\rm rms}$, respectively. We plot both relations in Fig.~\ref{fig:q_tau_c}, and we find $\tauc\sim 1/q$, which is in good agreement with the findings of \citet{2015MNRAS.446.2102N}, who derived this from velocity auto-correlation functions. We moreover find, $u_{\rm rms} \sim q^{3/2}$, and together this is equivalent to the $\eta(z\!=\!0)\sim q^2$ dependence shown in Fig.~\ref{fig:q_eta_yy}.  Since we obtain $\tauc$ from the assumed relation~(\ref{eq:eta}), the two are not independent, of course. We hence independently derive a scaling of $\tauc$ from the auto-correlation function of the kinetic helicity $h_{\rm kin}\equiv \overline{\U'\!\cdot(\nabla\tms\U')}$, which we use as a rough proxy for velocity correlations. This is plotted as a solid line in Fig.~\ref{fig:q_tau_c}, and exhibits a similar scaling behavior.

\subsection{Cycle Period as a Function of Shear Rate} 
\label{sec:cycle}

In the shear-dominated limit, $|\alpha|\,k_z (q\Omega)^{-1}\ll 1$, the cycle frequency of the $\alpha\Omega$ dynamo can be approximated as:
\begin{equation}
  \omega_{\rm cyc} \simeq \left|
  \frac{1}{2}\,\alpha_{yy}\, q\Omega\,k_z \right|^{1/2}\,,
  \label{eq:cyc}
\end{equation}
see, e.g., eqn.~(6.40) in \citet{2005PhR...417....1B}. Expressed in terms of the cycle period, $P_{\rm cyc}\equiv 2\pi/\omega_{\rm cyc}$, and including the scaling relation for $\alpha$ from Fig.~\ref{fig:q_dynamo_yy}, we predict the cycle period of the butterfly diagram to scale as\footnote{In view of the eleven-year solar cycle, the unique numerical constant of eleven `years' in equation~(\ref{eq:cyc_fit}) is a rather ironic coincidence.}
\begin{equation}
  P_{\rm cyc} \simeq \ \frac{2\pi}{\Omega} \ 
  \left( \pi\,q\; \frac{\alpha_0\,q}{H\Omega}\;
                       \frac{k_z H}{2\pi}\right)^{-1/2}
  \,\simeq\ \frac{11.0}{q}\ \frac{2\pi}{\Omega}\,,
  \label{eq:cyc_fit}
\end{equation}
where, in the last step, we have used a constant $\alpha_0 \simeq 0.008\,H\Omega$, corresponding to the dashed line labeled `$\sim\!q$', in the inset of Fig.~\ref{fig:q_dynamo_yy}, and have arbitrarily assumed a wavenumber $k_z\simeq 0.33$ in units of $2\pi/H$.

It is not straightforward to extract a meaningful wavenumber from the vertical profiles of $B_x$ and $B_y$, but since expression~(\ref{eq:cyc}) also appears in the real part of the eigenvalue,
\begin{equation}
  \mathcal{R}e\,\lambda_{\pm} \simeq -\eta_{\rm yy}\,k^2
  \,\pm\; \left|\frac{1}{2}\,\alpha_{yy}\, q\Omega\,k_z \right|^{1/2}\,,
  \label{eq:re_lambda}
\end{equation}
cf. eqn.~(6.39) in \citet{2005PhR...417....1B}, there is an alternative way to obtain $k_z$. Using our values for $\eta_{yy}$ (see Tab.~\ref{tab:results}), and assuming that the dynamo is one-dimensional ($k^2\equiv k_z^2$), and \emph{marginally} excited ($\mathcal{R}e\,\lambda=0$), we can use this to determine an independent estimate for the wavenumber, which we find to be $k_z\simeq 0.5$, using peak values for $\eta_{yy}$. Consistent with the scalings we find for $\alpha$ and $\eta$, and the way these quantities appear in equation~(\ref{eq:re_lambda}), this value is largely independent of the shear rate, $q$.

\begin{figure}
  \center\includegraphics[width=0.95\columnwidth]{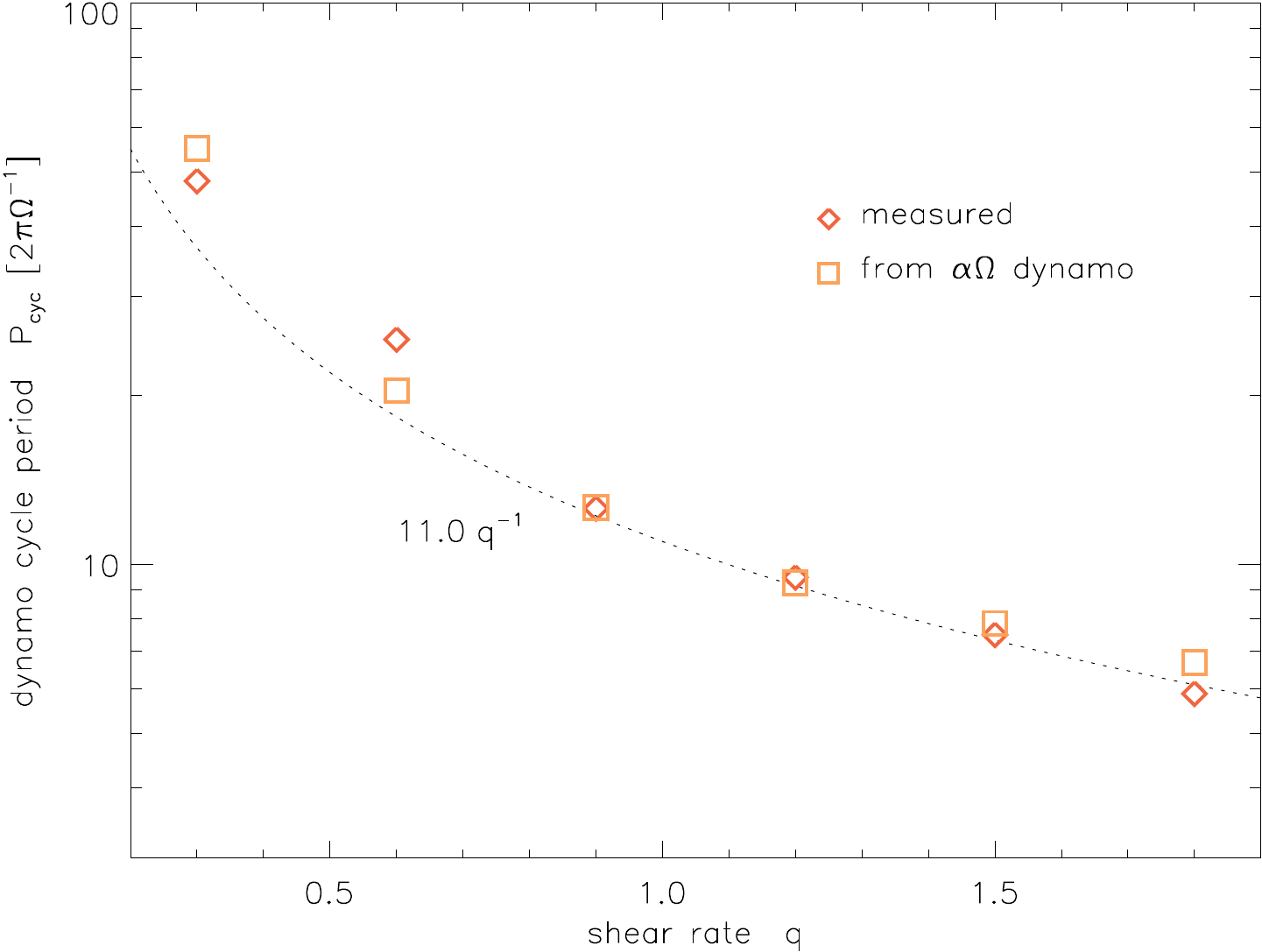}
  \caption{Dynamo cycle period, $P_{\rm cyc}$ as a function of the shear parameter. The periods measured via equation~(\ref{eq:fit_acf}) are shown as diamonds, and match well the prediction from equation~(\ref{eq:cyc}), with peak values for $\alpha_{yy}$ taken from Table~\ref{tab:results}, and $k_z=0.33$. The fitted relation~(\ref{eq:cyc_fit}) is shown as a dotted line.}
  \label{fig:q_dynamo_cyc}
\end{figure}

To reliably extract the cycle period from the space-time butterfly diagram observed in the various simulation runs, we apply the following procedure. We take $\bar{B}_y(z,t)$ profiles, that have been obtained by averaging in the $x$ and $y$ directions, and normalize them with the maximum at each given time $t$. This is to artificially enhance the contrast of the cyclic pattern. We then compute the autocorrelation function of $\bar{B}_y(t,z=z_i)$ at each height with $|z_i|>2.25\,H$. This exploits the fact that the cycle is seen most pronounced in the upper disk layers. Since $\bar{B}_y(t)$ is periodic in time, so is the ACF, $\hat{B}_y(\tau)$, and we apply a least-square fit to model the ACF as a mixture of an oscillation and a purely stochastic part, that is,
\begin{equation}
  \hat{B}_{y}(\tau) = \left[ (1-a) + a\,\cos(2\pi\,\omega_{\rm cyc}\,\tau)
    \right]\, {\rm e}^{-\tau/\tau_{\rm c}} \label{eq:fit_acf}\,,
\end{equation}
with the coefficient $a\in[0,1]$ being the relative amplitude of the oscillatory part. A function of this kind was used in \citet{2011MNRAS.415.3291G} for a different purpose. We find that this provides a useful method to extract the quasi-periodic cycle frequency $\omega_{\rm cyc}$, and we obtain cycle periods of $P_{\rm cyc}=\,$ 48.1, 25.2, 12.6, 9.5, 7.5 and 5.9 orbits for the models with $q=\,$ 0.3, 0.6, 0.9, 1.2, 1.5, and 1.8, respectively. These values are plotted together with the prediction~(\ref{eq:cyc}) and the derived scaling~(\ref{eq:cyc_fit}) in Fig.~\ref{fig:q_dynamo_cyc} as a function of the shear parameter. Given the crudeness of the approach, the agreement is striking and encourages further investigations into this direction. For instance, equation~(\ref{eq:cyc}) assumes a constant $\alpha$~coefficient and neglects any spatial variation. Moreover, we have taken the liberty to specify the wave number, $k_z$, of the dynamo mode \emph{ad hoc}, roughly based on the shape of the $\bar{B}_y(z)$ profile. 

Another reason for which our findings are surprising arises from the fact that we have used the values of the dominant \emph{positive} $\alpha$~effect (i.e., the squares in Fig.~\ref{fig:q_dynamo_yy}), and not the negative effect (triangles) near the disk midplane as suggested by \citet{1998tbha.conf...61B} to explain the `poleward' migration of the dynamo pattern.\footnote{Using the negative $\alpha$~effect did not provide a good fit with the cycle data, mostly because the amplitude is too low to match the order of magnitude with any reasonable choice of $k_z$. Note, however, that \citet{2002GApFD..96..319B} found the cycle period to be affected by non-locality in $\alpha$.}  This notion is further supported by our observation that the negative, buoyancy-related effect vanishes for $q\simgt 1.5$, whereas the migrating dynamo pattern persists for all studied values of the shear parameter. If the evidence obtained via the TF method holds up, this means that the explanation for the upward propagation of the field patterns will have to be revised.
This should however not be seen as a failure of the mean-field dynamo framework. After all, the notion that a positive (negative) $\alpha$~effect leads to equatorward (poleward) migration of field belts in the $\alpha\Omega$ dynamo, is based on the assumption of a constant \emph{isotropic} $\alpha$~tensor \citep[e.g.,][]{2005PhR...417....1B}. This is not the case for MRI-driven turbulence, where we observe a strong \emph{symmetric} contribution in the off-diagonal elements of both the $\alpha$ and $\eta$ tensors. This will mandate a thorough derivation of the mean-field dispersion relation, considering the mutual strength of the closure coefficients as well as their spatial variation.

\begin{figure}
  \center\includegraphics[width=\columnwidth,angle=0]{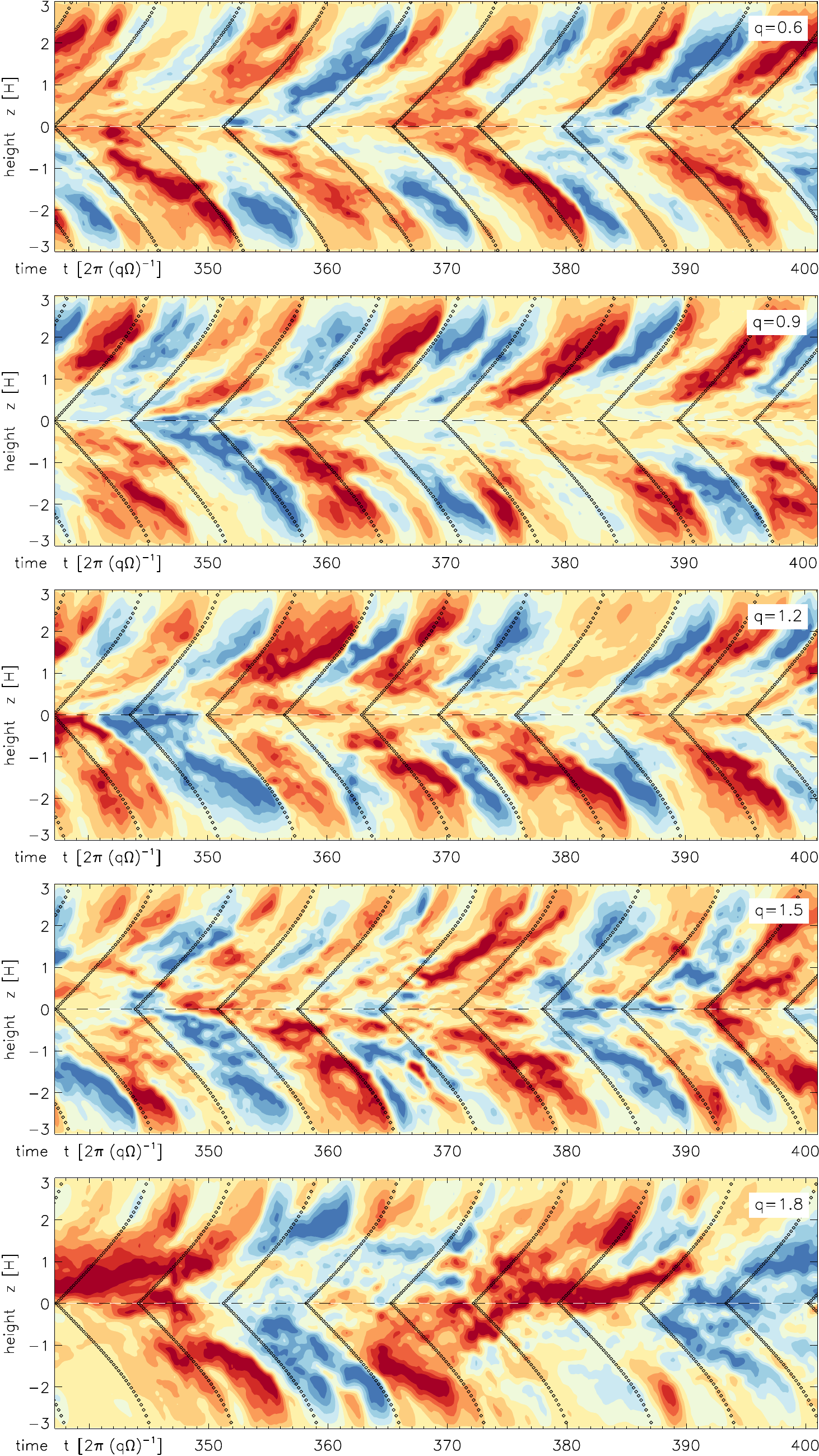}
  \caption{Azimuthal-field butterfly diagram for the models with $q=0.6$, 0.9, 1.2, 1.5, and 1.8 (top to bottom). Note that the time axis has been rescaled with $q$ to allow for a more direct comparison. The black lines illustrate the the cycle period and pattern speed, $c\equiv \omega_{\rm cyc}\,k_z^{-1}$ (albeit with an arbitrarily chosen propagation direction), of the $\alpha\Omega$~dynamo wave, and also include the advective contribution $\bar{u}_z$, which becomes apparent for $|z|\simgt 2\,H$.}
  \label{fig:propa}
\end{figure}

Figure~\ref{fig:propa} shows the prediction of the pattern speed from the most simple $\alpha\Omega$ model with a constant, isotropic, $\alpha$~effect. Because of the ``wrong'' sign of our $\alpha$, the propagation direction has been chosen arbitrarily. Note that the strict rule about downward (upward) propagation for positive (negative) $\alpha$ was in fact derived from a \emph{homogeneous} model. In an inhomogeneous situation such as ours, the dynamo can be affected by other factors -- see, for instance, \citet{2009MNRAS.398.1414B}, who found a reversal of the direction for wave propagation for different vertical boundary conditions in a stratified dynamo model. On top of the effect that is caused by the advective contribution, $\bar{u}_z$ (responsible for the curvature of the characteristics shown in Fig.~\ref{fig:propa}), there is a slight tendency towards faster propagation at larger $z$. This illustrates that a more detailed dynamo model is required to explain the field evolution in more detail \citep[also cf. figure~6 in][]{2010MNRAS.405...41G}.

\subsection{Net-vertical field dependence} 
\label{sec:nvf-dep}

\begin{figure}
  \center\includegraphics[width=0.95\columnwidth]{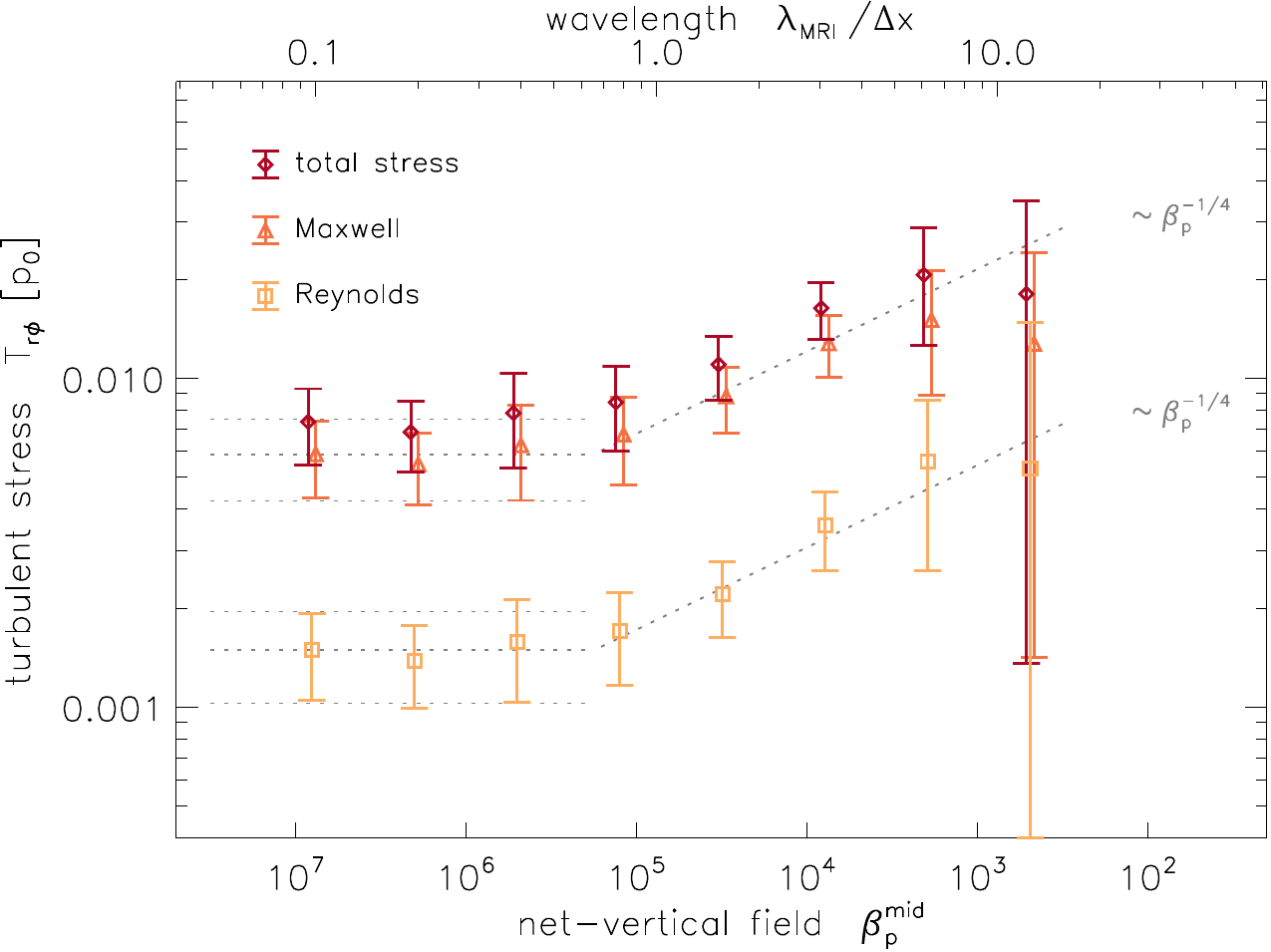}
  \caption{Net-vertical field dependence of the turbulent Reynolds stress (squares), Maxwell stress (triangles), and their sum (diamonds) for fixed $q=1.5$. The horizontal dotted lines indicate ZNF values, while the slanted lines follow a $\sim\betap^{-1/4}$ dependence. For the strongest field case, the disk is effectively disrupted by the vigorous MRI turbulence. The stress ratio is found to be independent of the NVF strength.}
  \label{fig:nvf_stress}
\end{figure}

Guided by the endeavor of developing a mean-field framework relevant to turbulent transport of angular momentum in accretion disks, we now turn our attention to the influence of varying levels of net-vertical magnetic flux. 

It has been understood very early that the amount of net-vertical field threading the disk has a direct influence on the accessible linear MRI modes even in the state of vigorously developed turbulence \citep{1995ApJ...440..742H}. The wavelength, $\lambda_{\rm MRI}$, of the fastest-growing linear mode is directly related to the Alfv{\'e}n speed corresponding to the vertical field component, that is
\begin{equation}
  \lambda_{\rm MRI} \equiv
  2\pi\sqrt{\frac{16}{15}}\,\frac{v_{{\rm A}z}}{\Omega}
  = 2\pi\sqrt{\frac{32}{15\,\betap^{\rm mid}}}\;H\,,
  \quad\textrm{for}\ q=\frac{3}{2}\,.
\end{equation}
For the case of unstratified MRI, \citet{2007ApJ...668L..51P} have found a linear scaling of the turbulent stresses with the most unstable wavelength\footnote {Note that the simulations considered in \citet{2007ApJ...668L..51P} correspond to those carried out by \citet{2004ApJ...605..321S}, who reported a scaling proportional to $\lambda_{\rm MRI}^{3/2}$. This steeper scaling is derived when the simulations with $v_{{\rm A} z}/(L_z\Omega)=6.25\times10^{-6}$ and $1\times10^{-4}$ are excluded in the fit (see Figure 2 in \citealt{2007ApJ...668L..51P}). Including
these results leads to the linear scaling previously reported by \citet{1995ApJ...440..742H,1996ApJ...464..690H}.},
$\lambda_{\rm MRI}$, and accordingly the amplitude of the net-vertical field $\bar{B}_z$. The authors, however, point out that this scaling is likely to change when accounting for the buoyant loss of magnetic flux, which is likely unavoidable in the stratified case.
In a previous paper that included a magnetically decoupled midplane in a poorly ionized, stratified protoplanetary disk \citep{2012MNRAS.422.1140G}, we found the Maxwell stress to scale $\propto \lambda_{\rm MRI}^{5/3}$, but this precluded the effect of turbulence near the midplane.

In Figure~\ref{fig:nvf_stress}, we plot the dependence of the turbulent stresses on the net-vertical field strength. The influence of the NVF becomes noticeable at $\betap^{\rm mid}\simeq 10^5$, and the additional contribution\footnote{This interpretation tacitly assumes that the NVF case is bounded from below by a ZNF self-sustaining state, whose convergence with numerical resolution is however still under debate -- even in stratified shearing-boxes that support mean-field dynamos \citep{2010ApJ...713...52D,2014ApJ...787L..13B}.} to the turbulent stresses roughly scales $\sim\betap^{-1/4}$. The break \citep{2007ApJ...668L..51P} seen in Fig.~\ref{fig:nvf_stress} furthermore appears to coincide with the ability to resolve the MRI in the disk body (see the upper axis of Fig.~\ref{fig:nvf_stress}, where we show the wavelength, $\lambda_{\rm MRI}$, of the fastest-growing linear mode in units of the radial grid spacing, $\Delta x$). We remark that, because $\betap<\betap^{\rm mid}$, this resolution indicator provides a lower bound in the stratified disk. As such, the transition point between ZNF and NVF may shift further towards the left if going to higher resolution, but attacking this question is beyond the scope of the current paper \citep[however see Sect.~3.3 in][for a potential explanation]{2010ApJ...712.1241S}. The slope of $\sim\betap^{-1/4}$ seen in Fig.~\ref{fig:nvf_stress} is considerably less steep than for unstratified simulations, and we attribute this potentially to the importance of buoyant flux loss and the modified eigenmodes for stratified case \citep{2010MNRAS.406..848L,VGSBpaper}. The observed flux-stress relation is qualitatively similar to the one seen for intermediate disk layers\footnote{That is, at $z=\pm 2H$, depicted by green circles and blue dots, respectively.} in Figure~2 of \citet{2010ApJ...712.1241S}, who performed global ZNF simulations and analyzed whether small local patches of these were compatible with the flux-stress relation obtained in NVF box simulations.

In our model with the strongest net flux, we observe noticeable disruption of the disk. This is in agreement with similar simulations by \citet{2013ApJ...767...30B}, who found magnetically dominated disks (i.e., $\betap<1$ throughout the domain) for models with initial $\betap^{\rm mid}\le 10^3$. In our models, the strong turbulence goes along with severe mass loss through the upper boundaries (which we artificially replenish by rescaling the mass density in the bulk of the disk). The mass loss rate is on the order of $M_0\,(2\pi)^{-1}\Omega$, where $M_0$ is the mass initially contained in the simulation. Because of the strong outflow, simulations with taller vertical domains than $\pm 3H$, as used here, should be performed \citep[cf.][]{2013A&A...552A..71F}.

\begin{figure}
  \center\includegraphics[width=0.95\columnwidth]{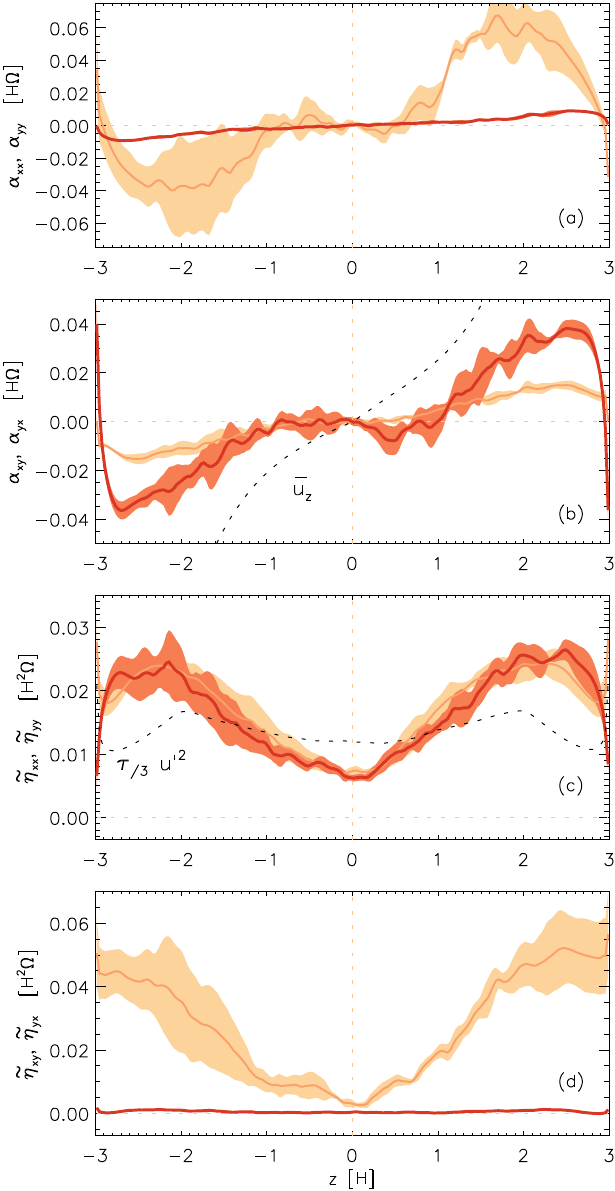}
  \caption{Same as Fig.~\ref{fig:dyn_q12}, but for the run with $q=1.5$, and $\bar{B}_z=0.32$, corresponding to a midplane $\betap\simeq 7800$.}
  \label{fig:dyn_nvf032}
\end{figure}

Keeping in mind these limitations of our current model, we nevertheless proceed with our investigation about how the NVF potentially affects the mean-field dynamo. In Fig.~\ref{fig:dyn_nvf032}, we show a representative example of dynamo profiles for a moderately strong net-vertical flux case. In comparison with Fig.~\ref{fig:dyn_q12}, for the ZNF case, we observe a significantly enhanced $\alpha_{xx}$~effect, whereas $\alpha_{yy}$ appears to be largely unaffected. Because of the dominant role of the $q\Omega\,\bar{B}_x$ term in the mean-field induction equation, even $\alpha_{xx}\simeq 0.1$ still provides only a relatively weak contribution to the overall $\alpha^2 \Omega$~dynamo. The coefficients $\alpha_{xy}$, and $\alpha_{yx}$ are less similar, and the latter also shows stronger fluctuations, as indicated by the shaded area. A further difference in the presence of a net-vertical field, is the disk wind, $\bar{u}_z$ that already rises steeply near the disk midplane.

Unlike in Fig.~\ref{fig:dyn_q12}, the turbulent diffusivities $\eta_{xx}$, and $\eta_{yy}$ now show a significant deviation from the kinematic result, prompting at the importance of magnetic fluctuations even for $|z|\simlt H$. Alternatively, the correlation time, $\tau$, could become dependent on the vertical position in the disk. Similarly to $\alpha_{xx}$, the off-diagonal component $\eta_{xy}$ is enhanced in the presence of a strong net-vertical field, even though, again the much smaller component $\eta_{yx}$ is the more interesting one in its relation to the dominant shear \citep[cf. appendix~B in][]{2005an....326..787b}. Despite the pronounced differences that we have just illustrated, the actual scaling behavior of the relevant dynamo coefficients with the NVF is found to be much less dramatic.

\begin{figure}
  \center\includegraphics[width=\columnwidth]{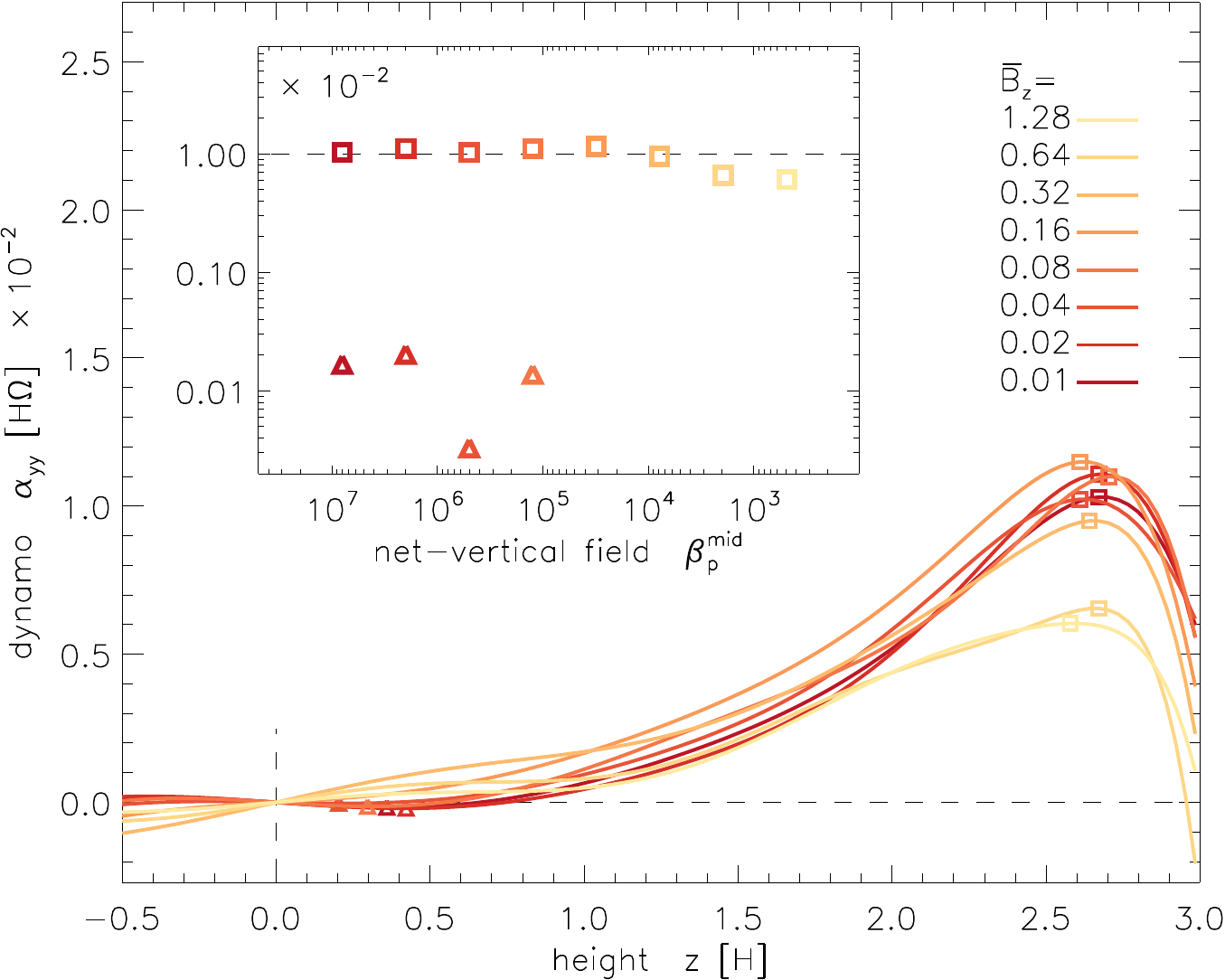}
  \caption{Vertical profiles (main plot), and net-vertical field dependence (inset) of the filtered $\alpha_{yy}(z)$ dynamo coefficient. As in the corresponding Fig.~\ref{fig:q_dynamo_yy}, amplitudes for the kinematic dynamo (squares), and the buoyancy-driven dynamo (triangles) have been inferred from peak values as indicated.}
  \label{fig:nvf_dynamo_yy}
\end{figure}

The dependence of $\alpha_{yy}(z)$ on $\betap^{\rm mid}$ is illustrated in Fig.\ref{fig:nvf_dynamo_yy}, where we plot their vertical profiles and infer the associated amplitudes. This is again done separately for the region near the midplane (triangles), and the upper disk layers (squares). The impact on the weak negative $\alpha$~effect is the most severe, and the effect vanishes for any appreciable amount of net-vertical flux (cf. Fig.~\ref{fig:nvf_stress}).  Compared with the shear-rate dependence shown in Fig.\ref{fig:q_dynamo_yy}, the variation of $\alpha_{yy}(z)$ in the upper disk with the NVF is minor, and only for very strong fields we observe a mild quenching. We speculate that this effect is related to the stronger outflow in this case, removing magnetic fluctuations before they can develop significant correlations with the fluctuating velocity. A quantitative assessment of this amounts to comparing the rotation timescale, $\sim\Omega^{-1}$, with the advective removal timescale, $\sim H\,\bar{u}_z^{-1}$. We remark that the observed trend is not compatible with what would be expected from an increasing effect of the advective helicity flux \citep{2009MNRAS.398.1414B}, that would likely be more pronounced in the case of a stronger NVF (and hence a stronger outflow). By removing small-scale current helicity from the system \citep[also see][]{2011ApJ...740...18O}, such a helicity flux would allow the dynamo to ``un-quench'' itself. It is however not straightforward to predict how such an effect would be reflected in the inferred TF coefficients that we present in Figure~\ref{fig:nvf_dynamo_yy}.

\begin{figure}
  \center\includegraphics[width=\columnwidth]{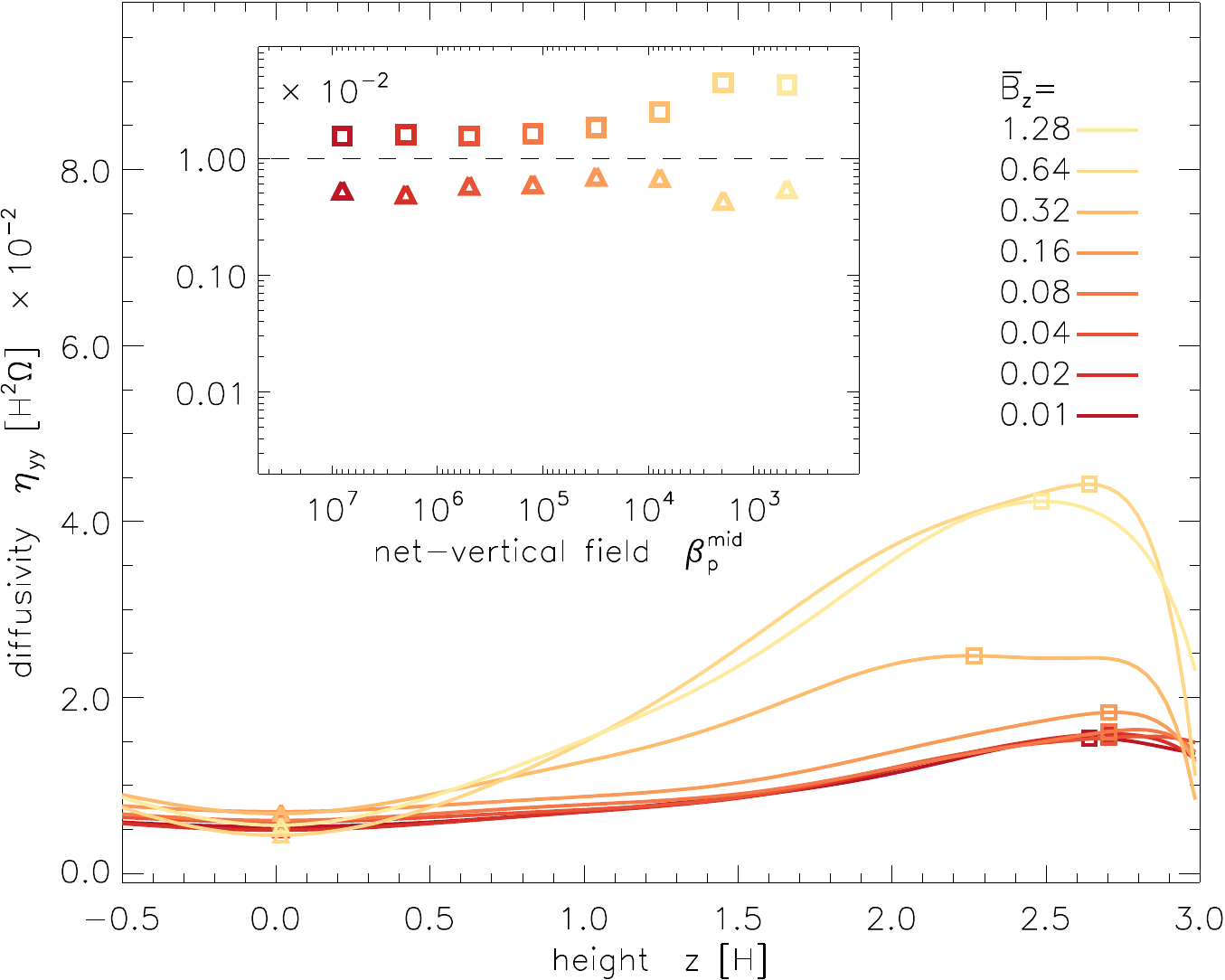}
  \caption{Same as Fig.~\ref{fig:nvf_dynamo_yy}, but for the diffusivity coefficient $\eta_{yy}(z)$. Peak values in the disk corona (squares) are enhanced at high NVF strength, while the midplane values appear to be quenched instead. Qualitatively similar scalings are found for the other tensor coefficients.}
  \label{fig:nvf_eta_yy}
\end{figure}

The influence on the turbulent diffusivity, when applying a NVF is shown in Fig.~\ref{fig:nvf_eta_yy}, where the midplane value, $\eta_{yy}(z=0)$, plotted with triangles, shows a similar quenching characteristic as seen for $\alpha_{yy}(z)$. Unlike the dynamo effect, the diffusivity displays an anti-quenching behavior in the upper disk layers (squares). Given the dominant role of magnetic tension forces ($\betap\simlt 1$) in the disk corona this is not necessarily surprising, and it fits together with the observation that the $\eta_{yy}(z)$ profiles deviate from the kinematic estimate in this region.

Based on the simple estimate~(\ref{eq:cyc}), which predicts the dynamo cycle frequency to depend on the square-root of the $\alpha$~effect, and looking at Fig.~\ref{fig:nvf_dynamo_yy}, we would not expect a strong variation of the observed cycle period with varying NVF. Unfortunately, the unambiguous determination of the cycles is hindered by the vigorous fluctuations in the simulations with significant net-vertical flux. A similar trend was reported by \citet{2013ApJ...767...30B} who found more irregular cycles for strong NVF, and state that the dynamo is ``fully quenched'' for strong background fields (i.e., $\betap^{\rm mid}=10^2$, where indeed no cycles are observed). With the rather moderate quenching seen in Fig.~\ref{fig:nvf_dynamo_yy}, we would however not interpret this as the absence of cyclic behavior (and consequently a mean-field dynamo) but merely as the cycles being dominated by other effects, such as linear MRI \citep[see discussion in ][]{2010MNRAS.405...41G} and parasitic modes \citep{1994ApJ...432..213G,2009MNRAS.394..715L,2009ApJ...698L..72P,2010ApJ...716.1012P}, which are moreover expected to develop on comparatively larger scales in the strong NVF regime. In contrast to the findings presented here, Figures~10 and 11 in \citet{2012MNRAS.422.1140G} show a pronounced dependence of the cycle period on the applied net flux. This was however for a partially ionized protoplanetary disk model, where the ability of the MRI to develop was severely limited by significant Ohmic dissipation in the disk body. As mentioned before, the flux removal timescale likely plays a role in the presence of the strong wind, and this is of course not reflected in equation~(\ref{eq:cyc}).

\subsection{Scale-dependent closure coefficients} 
\label{sec:non-local}

The functional dependence proposed in equation~(\ref{eq:closure}) assumes that the EMF at a given point in time and space can be modeled in terms of the mean magnetic field, and its derivatives, evaluated at the same point in time and space. This assumption can be relaxed so that the EMF is no longer local in space, according to
\begin{equation}
  \EMF_i(z) = \int \hat{\alpha}_{i\!j}(z,\zeta)\ \bar{B}_j(z-\zeta)
                \ -\ \hat{\eta}_{i\!j}(z,\zeta)\ \varepsilon_{\!jzl}\,\partial_z
                \bar{B}_l(z-\zeta)\; {\rm d}\zeta\,.
  \label{eq:closure_convolutions}
\end{equation}
Here, the tensor components $\hat{\alpha}_{i\!j}(\zeta)$ and $\hat{\eta}_{i\!j}(\zeta)$ play the role of convolution kernels that may additionally depend on the vertical coordinate $z$ -- reflecting the changing conditions with height in our stratified model. The spatial scale $\zeta$ over which they decay provides a measure of how the EMF at a given location depends on its surroundings. Note that equation~(\ref{eq:closure_convolutions}) recovers (\ref{eq:closure}) if $\hat{\alpha}_{i\!j}(z,\zeta) = \alpha_{i\!j}(z)\, \delta(\zeta)$ and $\hat{\eta}_{i\!j}(z,\zeta) = \eta_{i\!j}(z)\, \delta(\zeta)$, where $\delta(\zeta)$ is Dirac's delta function. In the following, we will omit the $z$~dependence in our notation for clarity. Because the EMF is a convolution in real space, its Fourier transform, denoted with a tilde, is simply
\begin{equation}
  \tilde{\EMF}_i(k_z) = \tilde{\alpha}_{i\!j}(k_z)\ \widetilde{\bar{B}_{\!j}}(k_z) \ -\ \tilde{\eta}_{i\!j}(k_z)\ {\rm i} k_z \,\varepsilon_{\!jzl} \,\widetilde{\bar{B}_l}(k_z) \,,
  \label{eq:closure_fourier}
\end{equation}
which provides the context for considering scale-dependent closure coefficients and illustrates the close relation between scale dependence and non-locality (with conjugate variables $\zeta\leftrightarrow k_z$).

\citet{2002GApFD..96..319B} first mentioned the possibility of a scale-dependent, or non-local, $\alpha$~effect in the context of stratified MRI (cf. their Figures 10-12). Even though their approach was affected by poor statistics,\footnote{Lacking the additional bits of information that the TF method so elegantly provides, their inverse problem was overconstrained and they had to resort to making assumptions about the tensorial structure of the $\eta$ coefficient.} the overall assumption appeared very justifiable. A more systematic approach to a non-local formulation for the EMF via scale-dependent $\tilde{\alpha}_{i\!j}(k_z)$ and $\tilde{\eta}_{i\!j}(k_z)$ coefficients can be naturally aided by employing scale-dependent test-fields. When the test fields appear as an inhomogeneity with a specific vertical wavenumber $k_z^{\rm TF}$ \citep{2008an....329..725b} this allows us to characterize the response of the system as a function of this imposed scale. By surveying a number of scales we can measure $\tilde{\alpha}_{i\!j}(k_z^{\rm TF})$ and $\tilde{\eta}_{i\!j}(k_z^{\rm TF})$.

Based on a series of harmonic test fields, \citet{2008A&A...482..739B} first introduced a scale-dependent TF method, and determined the shape of the convolution kernel for isotropically forced homogeneous turbulence.  They showed that the Fourier transforms of the convolution kernels can be characterized by Lorentzians,
\begin{equation}
  \tilde{\alpha}(k_z) = \frac{\alpha_{0}}{1+(k_z/k_{\rm c})^2}\,,\qquad
  \tilde{\eta}(k_z)   = \frac{\eta_{0}}  {1+(k_z/k_{\rm c})^2}\,.
  \label{eq:kernel_fourier}
\end{equation}
This implies that the convolution kernels in real space are decaying exponentials,
\begin{equation}
  \alpha(\zeta) = \frac{\alpha_{0}}{2} \exp(-k_{\rm c}\,|\zeta|\,)\,,\qquad
  \eta (\zeta)  = \frac{\eta_{0}  }{2} \exp(-k_{\rm c}\,|\zeta|\,)\,,
  \label{eq:kernel_real_space}
\end{equation}
where the characteristic wavenumber, $k_{\rm c}$, can in principle take individual values for $\alpha$ and $\eta$. In fact, \citet{2008A&A...482..739B} found the characteristic scale over which the diffusivity $\eta$ falls off to be about a factor of two smaller than for $\alpha$, that is, $k^{(\eta)}_{\rm c}\simeq 2\,k^{(\alpha)}_{\rm c}$.

In what follows, we extend this approach and apply it to stratified MRI turbulence. While it is in principle possible to obtain all wavenumber components simultaneously, we have chosen to run a separate model for each wavenumber $k_z^{\rm TF}$. Unlike for the homogeneous turbulence studied by \citet{2008A&A...482..739B}, our coefficients $\alpha(z,k_z^{\rm TF})$, and $\eta(z,k_z^{\rm TF})$ are simultaneously functions of the position $z$, and the wavenumber $k_z^{\rm TF}$. The latter, however, only enters via the prescribed test-field inhomogeneity (as indicated by the label `TF') in the evolution equation for the passively evolved magnetic fluctuations. While this may appear somewhat confusing, we point out that the variables $z$, and $k_z^{\rm TF}$ are not complementary in the usual sense, that is, when applying a Fourier transform to some function $f(z)$ to obtain $\tilde{f}(k_z)$.

\begin{figure}
  \center\includegraphics[width=\columnwidth]{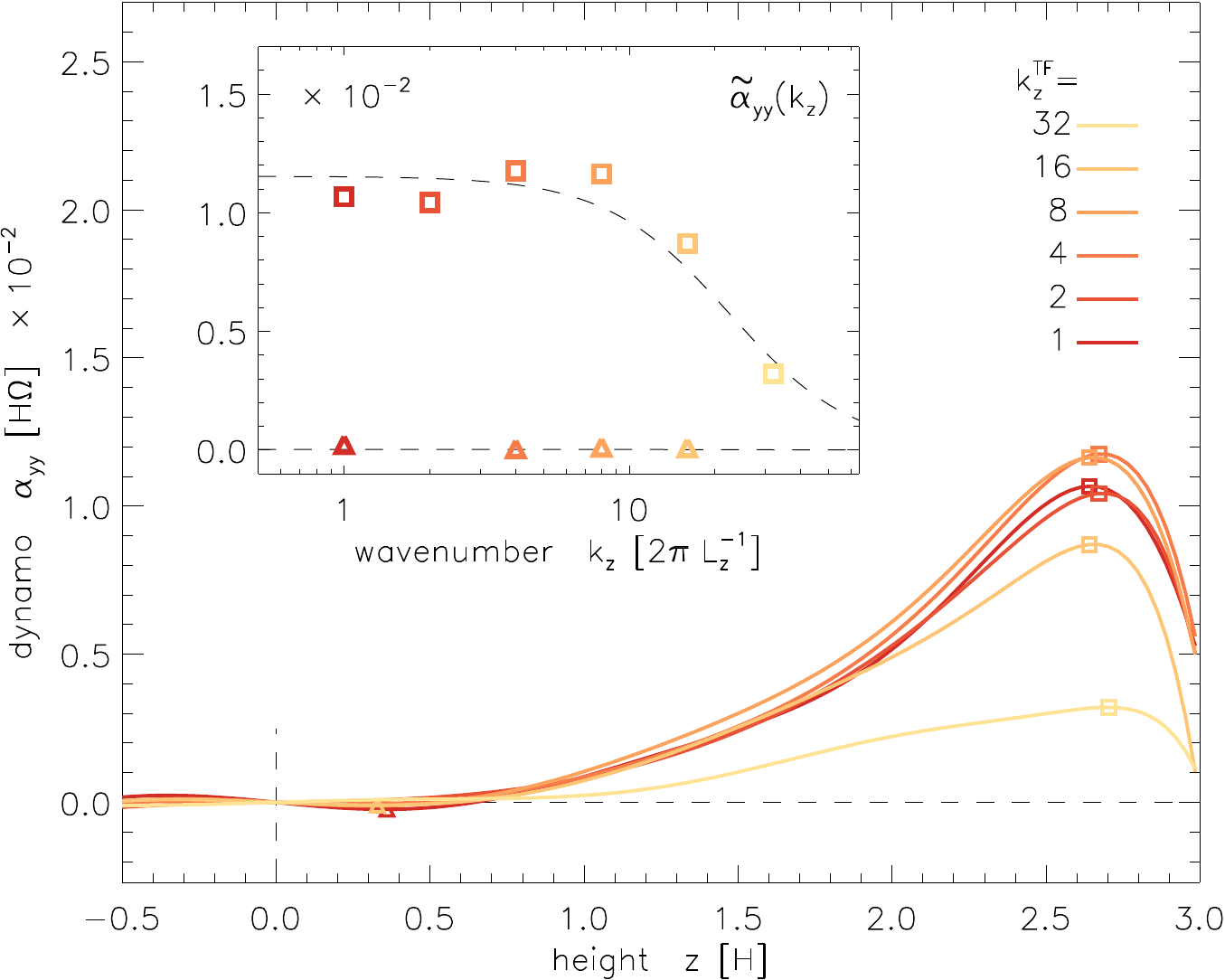}
  \caption{Vertical profiles (main plot), and wavenumber dependence (inset) of the $\alpha_{yy}(z)$ dynamo coefficient. As in the corresponding Fig.~\ref{fig:q_dynamo_yy}, amplitudes have been obtained for the kinematic dynamo (squares), and the buoyancy-driven dynamo (triangles). While the former scale similar to a Lorentzian (dashed) with $k_{\rm c}=3.7$, no clear scaling is seen for the latter.}
  \label{fig:k_dynamo_yy}
\end{figure}

Accordingly, in Fig.~\ref{fig:k_dynamo_yy} the ordinates in the main plot and inset are just as independent as in all the previous plots, and it is meaningful to look at the $k_z$~dependence at various positions $z$. The picture for the weak negative $\alpha$~effect (triangles) is uncertain, and no clear trend is seen with wavenumber. The stronger, positive $\alpha$ effect is flat for low wavenumbers, and its spectral shape is compatible with a Lorentzian of width $k_{\rm c}^{(\alpha)}=3.7$ in units of $2\pi/H$ (note that the axis in Fig.~\ref{fig:k_dynamo_yy} is in units of $2\pi/L_z$).
\begin{figure}
  \center\includegraphics[width=\columnwidth]{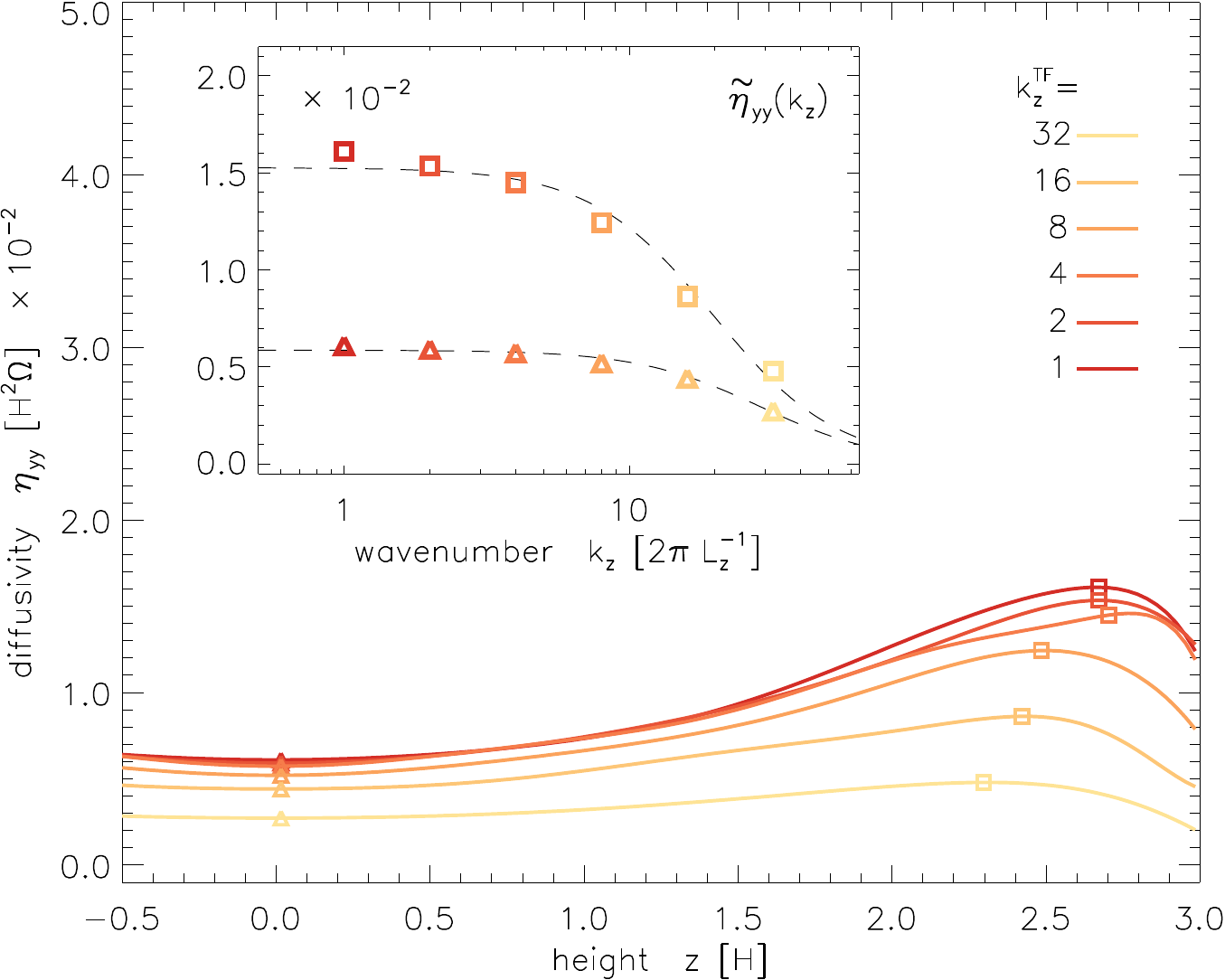}
  \caption{Same as Fig.~\ref{fig:k_dynamo_yy}, but for the diffusivity coefficient $\eta_{yy}(z)$. Peak values in the disk corona (squares) as well as the midplane values (triangles) are both compatible with Lorentzian shapes (dashed lines).}
  \label{fig:k_eta_yy}
\end{figure}
If we look at Fig.~\ref{fig:k_eta_yy}, a similar scaling is found for the turbulent diffusivity, $\eta_{yy}(z)$ -- both for the disk corona (squares) for which we find $k_{\rm c}^{(\eta)}=3.3$, and for the midplane values (triangles) for which we find $k_{\rm c}^{(\eta)}=4.8$. As we have mentioned already, \citet{2008A&A...482..739B} had found a wider kernel (by a factor of two) for $\alpha$, compared to $\eta$ in their simulations. Here we find the opposite trend with $k_{\rm c}^{(\alpha)}$ being roughly 15\% larger than $k_{\rm c}^{(\eta)}$ when comparing the results obtained at $z\simeq2.5\,H$. However, the value of $k_{\rm c}^{(\eta)}$ increases by about 50\% near the midplane, likely reflecting the shorter correlation length at higher density. Compared with the wavenumber of the large-scale dynamo mode, that we estimated as $k_z\simeq 0.33$ in Section~\ref{sec:cycle}, a value of $k_{\rm c}\simeq 3.3$ implies that our simulations have a scale-separation ratio of about ten. While this is sufficiently large to warrant a mean-field description in general, it may not be comfortably so -- hence making it worthwhile to consider non-local contributions to the EMF in addition.


\section{Summary and Discussion}
\label{sec:summary}

We continue our endeavor to build a unified framework for angular momentum transport in accretion disks, on one hand, and the redistribution and amplification (via a dynamo) of large-scale magnetic fields, on the other hand \citep{2010AN....331..101B}. In the present paper, we have studied various parameter dependencies of the turbulent transport coefficients. The main focus was on the shear rate, $q$, and the net-vertical field strength (quantified by the midplane plasma parameter $\betap$). Simultaneously, we have investigated how the closure coefficients for the classical $\alpha\Omega$~dynamo depend on these parameter variations.  The correspondence between these two aspects promises to increase our knowledge about the inner workings of accretion disk dynamos and ultimately about the mass transport within such disks. 

Following a complementary avenue, we have studied the scale-dependence of dynamo closure coefficients, greatly improving over a previous attempt by \citet{2002GApFD..96..319B}. Compared to their result, the TF method offers a much higher fidelity in determining the scale-dependence of the dynamo coefficients, and at the same time allows us to simultaneously obtain $z$-dependent profiles.

\subsection{Shear-rate dependence and dynamo cycles}

Our results on the shear-rate dependence are congruent with recent work by \citet{2015MNRAS.446.2102N}, who found that the total (kinetic plus magnetic) turbulent stress scales with the shear-to-vorticity ratio, $q/(2-q)$, as previously suggested by \citet{1996MNRAS.281L..21A}. As predicted by \citet{2006MNRAS.372..183P}, and as also found by \citet{2015MNRAS.446.2102N}, the Maxwell-to-Reynolds stress ratio scales $T_{R\phi}^{\rm Maxw}/T_{R\phi}^{\rm Reyn}\sim\,(4-q)/q$, and we find a value of $\simeq 4$ for Keplerian shear. 

As in previous work \citep{2010MNRAS.405...41G,2013ApJ...770..100G}, we apply the test-field method to obtain measurements for the mean-field $\alpha$~effect, and eddy diffusivity, $\eta$, which we find to significantly depend on the background shear-rate. The turbulent diffusivity is found to scale like $\eta \sim q^2$ near the disk midplane, whereas a scaling $\eta \sim q$ is found in the upper disk layers. The quadratic scaling can be explained by the dependence of the turbulent correlation time $\tau \sim 1/q$ \citep[also cf.][]{2015MNRAS.446.2102N}, and the scaling of the turbulent velocity dispersion $u_{\rm rms} \sim q^{3/2}$.

Motivated by the claim \citep{1995ApJ...446..741B} that the `poleward' migrating dynamo pattern requires a \emph{negative} $\alpha$~effect, we separately inspected the negative $\alpha_{yy}$ that we find near $z\simeq 0.5\,H$, and the positive $\alpha_{yy}$ measured in the disk corona ($z\simeq 2.5\,H$). The former effect was predicted by \citet{2000A&A...362..756R} to only exist in a narrow window of shear-rates, $q$. We indeed confirm this behavior by fitting a heuristic function $\;q^4\,(1.5-q)\;$. In contrast, the stronger, positive $\alpha$~effect was found to exist for all studied shear rates $q=0.3,\dots, 1.8$, with a linear scaling $\alpha_{yy}\sim q$. A linear scaling with shear has been found previously in a study of \emph{unstratified} turbulence subject to artificial forcing \citep{2008PhRvL.100r4501Y}. In their case, the dynamo growth rate was found to depend linearly on the shear, as explained by linear theory \citep{2011PhRvL.107y5004H}. We remark, however, that this effect is likely distinct from the present case, where the dynamo effect is probably enabled by stratification rather than shear alone \citep[see the discussion in][]{2013ApJ...762..127B}.

The above interpretation tacitly assumes that the observed scaling is a result of the \emph{production} mechanism. Since the measured $\alpha$~effect is likely in a non-linear, saturated regime, the scaling may equally well reflect the \emph{quenching} mechanism. This opens the possibility to test ideas related to shear-driven helicity fluxes \citep{2001ApJ...550..752V,2008an....329..725b,2009ApJ...696.1021V,2011ApJ...727...11H,2014ApJ...780..144V}, and their role in determining dynamo saturation. Regardless of its origin, the linear scaling, in any case, nicely explains the shear-rate dependence of the dynamo cycle period, which we successfully predict using the dispersion relation for the standard $\alpha\Omega$~dynamo. The striking agreement suggests that the positive $\alpha$~effect may be responsible, after all, for the characteristic ``butterfly'' cycles seen in stratified MRI simulations. More supporting evidence for this hypothesis comes from the observation that we do see cyclic patterns in all studied cases for $q$, whereas the negative $\alpha$~effect is found to vanish for $q\simgt 1.5$. A definite resolution of what effect determines the cycle period, and what effect sets the propagation direction of the dynamo pattern will have to await quantitative mean-field modeling -- taking into account the full tensorial structure of both the $\alpha$ and $\eta$ tensors.

\subsection{Dependence on net-vertical flux}

The inclusion of a significant net-vertical magnetic field affects the amplitude of the turbulent stresses -- yet not as dramatically as in unstratified simulations \citep[compare Fig.~\ref{fig:nvf_stress} with Fig.~2 in][]{2007ApJ...668L..51P}. For our simulation setup, we find a threshold value of $\betap \simeq 10^5$, expressed in the midplane plasma parameter. For values smaller than this (i.e., for stronger fields), we find an approximate scaling with \,$\betap^{-1/4}$. Inclusion of the NVF appears to have a varying effect on the various tensor components.  We observe some of the coefficients, such as $\alpha_{xx}$, and $\eta_{xy}$, to be significantly enhanced for stronger NVF. The key coefficient $\alpha_{yy}$ entering the $\alpha\Omega$~cycle, however, shows the opposite trend, and is weakly quenched for strong fields. A similar dichotomy applies to the turbulent diffusivity, represented by $\eta_{yy}$, which shows a weak quenching near the disk midplane, and is slightly anti-quenched in the upper disk layers. We interpret the observed characteristic to be caused by the faster removal timescale due to the stronger wind for higher NVF, such that correlations (caused primarily by the Coriolis effect) have shorter time to build-up in models with stronger NVF and hence a stronger wind. The advective helicity flux \citep{2009MNRAS.398.1414B} will of course also scale with the outflow velocity, offering an alternative potential explanation for the observed behavior.

\subsection{Scale separation and non-locality}

Motivated by the paper of \citet{2002GApFD..96..319B}, and with the technological advancement introduced by \citet*{2008A&A...482..739B}, we have obtained the scale-dependence of the $\alpha$~effect, and turbulent $\eta$, for stratified, isothermal MRI turbulence with unprecedented fidelity. \citet{2002GApFD..96..319B} for instance found $\eta_{xx}\simeq 10\times \eta_{yy}$ (see their Figure~12). In contrast, we find both components of comparable magnitude. This is likely explained by their assumption of negligible off-diagonal contributions during the inversion process, which is however not warranted in view of our TF results that show $\eta_{xy}\simeq \eta_{xx} \simeq \eta_{yy}$. Without this restrictive assumption, \citet{2002GApFD..96..319B} find unphysical negative diffusivities for the coefficient affecting the radial magnetic field component (see their figure 10), indicating an inherent problem with their original approach.

Using the spectral TF method, for the $\alpha_{yy}$ tensor element, we find a wavenumber dependence $\sim (1\!+\!k_z^2/k_{\rm c}^2)^{-1}$ with a characteristic wavenumber of $k_{\rm c}^{(\alpha)}=3.7$ (in units of $2\pi/H$) at a disk height of $z\simeq 2.75\,H$. The Lorentzian shape in $k$~space has first been identified by \citet{2008A&A...482..739B} and \citet{2009A&A...495....1M}, and also appears in passive scalar diffusion \citep[see Sect.~III.C of][]{2010PhRvE..82a6304M}.

For the eddy diffusivity $\eta(k)$, we find a similar dependence, but with $k_{\rm c}^{(\eta)}=3.3$ (in the magnetically dominated corona), and $4.8$ (near the disk midplane). This illustrates that the non-locality in the relation between the mean magnetic field and the EMF manifests itself differently in different regions of the disk, and this is the first time that both the vertical functional dependence \emph{and} the scale-dependence of these effects has been derived simultaneously, adding credibility to their mutual consistency \citep[see the concerns in][]{2002GApFD..96..319B}. With the dominant wavenumber of the mean-field dynamo estimated at $k_z\simeq 0.33$, and $k_{\rm c}\simeq 3.5$, we infer a scale separation close to one order of magnitude, which comfortably merits a mean-field description -- yet not without concerns that non-local effects may potentially affect the properties of the emerging dynamo solution.


\section{Conclusions}
\label{sec:concl}

Its ability to quantitatively explain the period of the observed cyclic field patterns (as a function of shear rate) encourages to further explore the $\alpha\Omega$~mechanism as a key player in the long-term evolution and dynamics of magnetized accretion disks. The unambiguous determination of the inherent scale-dependence in the dynamo tensors furthermore instigates a new investigation into more realistic non-local mean-field models. \citet{2002GApFD..96..319B} have already established that such an approach does affect properties of the butterfly diagram in simpler settings.

A number of issues need to be understood along the way, such as the largely unknown (anti-) quenching behavior of the mean-field coefficients in the case of a dynamically important Lorentz force \citep[even including the possibility of spontaneous symmetry breaking;][]{2011PhRvE..84b5403C}. A stronger $\alpha$~effect for stronger mean fields was, for instance, argued to potentially explain the steep scaling of the dynamo cycle period, $\omega_{\rm cyc}$, with the rotation frequency observed in stellar dynamos \citep{2011A&A...534A..46C}.

Alternatively, one can strive to include the effects of the Lorentz force explicitly. The latter route would require to extend the mean-field approach to the momentum equation -- something that can be regarded as mandatory if one wants to include the modeling of turbulent stresses \citep{2003MNRAS.340..969O,2006PhRvL..97v1103P}. Additionally, dynamical constraints from the conservation of magnetic helicity \citep[see][for a recent account]{2012ApJ...748...51H} may govern at least part of the dynamics \citep[as witnessed by the dynamically quenched mean-field model of][]{2010MNRAS.405...41G}.

Before global simulations were feasible, a large body of work employed the local shearing-box approximation \citep[to mention two influential ones]{1995ApJ...440..742H,1995ApJ...446..741B} to obtain bulk properties of MRI turbulence. In this paper, we also heavily rely on the shearing-box as a workhorse for studying differentially rotating turbulence. The renewed realization \citep{2009ApJ...691L..49S} of the importance of outflows \emph{\`{a} la} \citet{1982MNRAS.199..883B} challenges the usefulness of the local approximation, where some properties of the outflow are affected by spurious box-size effects \citep{2013A&A...552A..71F}. While this may also partly affect our results with strong net-vertical fields, we nevertheless argue that only \emph{combining} local and global models can ultimately solve the puzzle of understanding turbulent accretion disks.

All in all, the approach presented in this paper appears to be a sensible first step in a program to build a mean-field framework for turbulent flows that will enable to follow the secular evolution of global disk dynamics. The fact that the properties of the turbulent dynamics, as well as the butterfly pattern, seen in local simulations are also observed in global simulations \citep{2012ApJ...744..144F,2014ApJ...784..169J} lends support to this approach.


\section*{Acknowledgments}

We thank Axel Brandenburg and Gopakumar Mohandas for useful discussions and comments on an earlier draft of the paper. We are also thankful to Chi-kwan Chan, Eric Blackman, and Farrukh Nauman for
useful conversations. This work used the \NIII code developed by Udo Ziegler at the Leibniz Institute for Astrophysics (AIP). We acknowledge that the results of this research have been achieved using the PRACE-3IP project (FP7 RI-312763) resource \texttt{Fionn} based in Ireland at the Irish Centre for High-End Computing (ICHEC). This work was supported by a research grant (VKR023406) from the Villum Foundation. The research leading to these results has received funding from the Danish Council for Independent Research (DFF) and FP7 Marie Curie Actions -- COFUND under the grant-ID: DFF -- 1325-00111, as well as the European Research Council under the European Union's Seventh Framework Programme (FP/2007-2013) under ERC grant agreement 306614. MEP acknowledges support from the Villum Foundation's Young Investigator Programme.


\end{document}